\documentclass[aps,prd,twocolumn,superscriptaddress,nofootinbib,showpacs,showkeys]{revtex4-1}

\usepackage{graphicx,color}
\usepackage{enumerate}
\usepackage{amsmath,amssymb,amsthm}
\usepackage[paperwidth=210mm,paperheight=297mm,centering,hmargin=20mm,vmargin=20mm]{geometry}
\usepackage{float}
\usepackage{mathrsfs}
\usepackage{hyperref}
\usepackage{physics}
\usepackage[T1]{fontenc}
\usepackage{slashed}
\usepackage[usenames,dvipsnames]{xcolor}
    \definecolor{officegreen}{rgb}{0.0, 0.5, 0.0}

\usepackage{accents}
\newcommand*{\dt}[1]{%
  \accentset{\mbox{\large\bfseries .}}{#1}} 
\newcommand*{\ddt}[1]{%
  \accentset{\mbox{\large\bfseries .\hspace{0ex}.}}{#1}}

\renewcommand{\i}{\text{i}}
\renewcommand{\d}{\text{d}}
\newcommand{\e}{\text{e}}

\newcommand{\be}{\begin{equation}}
\newcommand{\ee}{\end{equation}}
\newcommand{\beq}{\begin{eqnarray}}
\newcommand{\eeq}{\end{eqnarray}}

\date{\today}

\begin{document}
\title{Entanglement entropy at critical points of classical evolution in oscillatory and exotic singularity multiverse models}

\author{Adam Balcerzak}
\email{Adam.Balcerzak@usz.edu.pl}
\affiliation{Institute of Physics, University of Szczecin, Wielkopolska 15, 70-451 Szczecin, Poland}

\author{Samuel Barroso-Bellido}
\email{Samuel.Barroso-Bellido@usz.edu.pl}
\affiliation{Institute of Physics, University of Szczecin, Wielkopolska 15, 70-451 Szczecin, Poland}

\author{Mariusz P. D\c{a}browski}
\email{Mariusz.Dabrowski@usz.edu.pl}
\affiliation{Institute of Physics, University of Szczecin, Wielkopolska 15, 70-451 Szczecin, Poland}
\affiliation{National Centre for Nuclear Research, Andrzeja So{\l}tana 7, 05-400 Otwock, Poland}
\affiliation{Copernicus Center for Interdisciplinary Studies, Szczepa\'nska 1/5, 31-011 Krak\'ow, Poland}

\author{Salvador Robles-P\'{e}rez}
\affiliation{Canadian Quantum Research Center, 204-3002 32 Ave Vernon, BC V1T 2L7,  Canada}
\affiliation{Estaci\'{o}n Ecol\'{o}gica de Biocosmolog\'{\i}a, Pedro de Alvarado, 14, 06411 Medell\'{\i}n, Spain.}

\begin{abstract}
Using the 3rd quantization formalism we study the quantum entanglement of universes created in pairs within the framework of standard homogeneous and isotropic cosmology. In particular, we investigate entanglement quantities (entropy, temperature) around maxima, minima and inflection points of the classical evolution. The novelty from previous works is that we show how the entanglement changes in an extended minisuperspace parameterised by the scale factor and additionally, by the massless scalar field. We study the entanglement quantities for the universes which classically exhibit Big-Bang and other than Big-Bang (exotic) singularities such as Big-Brake, Big-Freeze, Big-Separation, and Little-Rip. While taking into account the scalar field, we find that the entanglement entropy is finite at the Big-Bang singularity and diverges at maxima or minima of expansion. As for the exotic singularity models we find that the entanglement entropy or the temperature in all the critical points and singularities
is either finite or infinite, but it never vanishes. This shows that each of the universes of a pair is entangled to a degree parametrized by the entanglement quantities which measure the quantumness of the system. Apart from the von Neumann entanglement entropy, we also check the behaviour of the the Tsallis and the Renyi entanglement entropies, and find that they behave similarly as the meters of the quantumness. Finally, we find that the best-fit relation between the entanglement entropy and the Hubble parameter (which classically marks special points of the universe evolution) is of the logarithmic shape, and not polynomial, as one could initially expect. 

\end{abstract}

\maketitle

\section{Introduction}


Since its beginning \cite{DeWitt1}, the canonical quantum cosmology suffers from lack of explanations, like, for example, what happens with the arrow of time or finding a unique solution for the wave function of the universe or the wave function interpretation (see for example \cite{KieferBook,PhysRevD.39.1116}).

There is, however, the formalism of the third quantization \cite{3rdQ} which is quite natural, if we believe in the existence of a wave function of the universe $\Psi$. This formalism interprets universes as particles, i.e., the perturbations around the vacuum of the virtual universes, where the most natural way to create them is in pairs as it is the case in quantum field theory. One universe plays the role of a particle and the other (anti-universe) of an antiparticle. In the simplest case, we  associate wave functions to those universes as complex scalar functions, and therefore, each of them is described by complex conjugated wave functions corresponding to particles and antiparticles.

The validity of the formalism is still under debate since there are not any experimental results to confirm it yet. Here, we present a recent formalism to analyse the quantumness of a pair of universes using their entanglement quantities, which could shed light on some aspects of quantum cosmology such as for instance the question if the quantumness is restricted to the initial singularity of the universe or takes place throughout the whole evolution of the universe. There have been some studies during last decades trying to test the existence of the multiverse as the collection of the universes (see for example \cite{1,2,2.5,3,4,5}). They considered the phenomenon of quantum entanglement between universes within the multiverse as essential for the task. The point was to calculate the entanglement entropy explicitly and to draw attention into some particularities of this approach.

In fact, the underlying Wheeler-DeWitt equation of canonical quantum cosmology is nothing else than the Hamiltonian constraint $H\Psi=0$ of the minisuperspace which we work with. It describes the wave function of each individual universe in a set of universes -- the multiverse. However, the application of the quantum theory to the universe as a whole challenges some of its fundamentals and forces us to re-think them. How to interpret the wave function $\Psi$ is clearly one of those fundamentals.

Let us point out that there are two main interpretations of $\Psi$ in the quantum theory. The first one, the more 'wave-like' interpretation of Quantum Mechanics, where the wave function represents the state of a system, possibly with an internal structure (let us think on an hydrogen atom), and whose quantum state can be described in terms of the elements of some basis of the corresponding Hilbert space. And the second one, the more 'particle-like' interpretation of the Quantum Field Theory, which describes a field whose quantum state can be interpreted in terms of particles that can in principle be measured separately. They are related, but there are subtle differences in the interpretation. For instance, in the former the basis state $| n \rangle$ does not represent any entity by itself but a specific \emph{level} (e.g., an energy level) in terms of which we can describe the state of the whole system. In the latter the basis state $| n\rangle$ do represent the number of more or less independent entities called \emph{particles}\footnote{Perhaps another way to characterise this dichotomy is to ask
what is exactly the entity: the field $\Psi$ or the particles.}. In fact, a possible way to characterise this dichotomy is to ask what is exactly the entity: the field $\Psi$ or the particles.

In quantum cosmology the former is the customary interpretation given to the wave function $\Psi$ \cite{Hartle1983, Vilenkin1988, Kiefer2013}, where it represents the state of the whole universe with its internal structure (spacetime $+$ matter fields). And the latter is exactly the interpretation adopted in the third quantisation formalism \cite{Caderni1984, McGuigan1988} (see also Ref. \cite{DeWitt1967}), where the universes are interpreted as particles moving in the superspace of spatial geometries and matter field configurations. Here, we are going to pose and analyse a phenomenon that is naturally contextualized in the third quantization formalism, the creation of a universe-antiuniverse pair, but from the point of view of their wave functions. This is justified because we do not want to study the 'many-particle' state of the multiverse but only the state and the quantum correlations of a couple of these newborn pair of universes.

At the beginning of the evolution, the global state of a pair of universes is taken as the ground state $\ket{00}$ for all quantum representations \cite{Interuniversal}. Once the evolution takes place, the vacuum state changes differently in different representations. For instance, the diagonal representation of the Hamiltonian $H$ is not invariant under the evolution of the universes which means that any instantaneous eigenstate $|N_0\rangle_d$ at a given moment of time spreads into the infinite set of the basis components, $\{| N\rangle_d \}_{N\in \mathbb{N}}$, of the diagonal representation at any later time. On the other hand, the invariant representation of the Hamiltonian $H$ is invariant under the evolution \cite{LewisRiesenfeld, KimLewis}. It means that a particular state $|N_0\rangle_i$ of the invariant representation remains the same state along the entire evolution. 
It seems therefore a very plausible boundary condition to be imposed that the universes of a universe-antiuniverse pair are in the ground state of the invariant representation\footnote{That implicitly assumes that the universe-antiuniverse pair is isolated, i.e. non-interacting with the rest of universe-antiuniverse pairs of the multiverse.} which implicitly assumes that the universe-antiuniverse pair is isolated, i.e. non-interacting with the rest of universe-antiuniverse pairs of the multiverse. As we have said, at the beginning of the evolution these two representations coincide, so the universes are also initially in the ground state of the instantaneous diagonal representation. However, as the universes evolve, their quantum states are represented at any later time by a superposition of excited states of the diagonal representation. It turns out that this state is a mixed state for which the entropy can be computed and is different from zero \cite{Interuniversal}.

As we know from quantum optics and quantum mechanics, the entropy of entanglement is a good measure of the \emph{quantumness} of a physical system, being this interpreted in terms of the quantum correlations existing between the states of the components. Thus, we can use the entropy of entanglement between a universe and its corresponding antiuniverse to study the degree of quantumness of the universe.  It was shown in Ref. \cite{Interuniversal} that the entanglement entropy is infinite at the maximum point of classical expansion of any universe in the multiverse. A possible weak point of these calculations could be the usage of the WKB approximation, where the wave function diverges precisely at the maximum. Then, in this paper we are not going to make any such an approximation and want to analyse the entanglement at any critical point during the evolution of the universe \textit{ in time}, that in quantum cosmology is supposed to be directly represented by the scale factor. Our procedure is going to be as exact as possible.

In section \ref{S2}, we deal with mathematical details of the third quantization: the existence of conjugated solutions of the Wheeler-DeWitt equation, their physical meaning, and the distribution of modes of the scalar field. In section \ref{E2dim} we analyse the exact entropy of entanglement of a universe with a massless scalar field as a second degree of freedom in the minisuperspace, in that way generalising and checking the results of \cite{Interuniversal}. In section \ref{EECP}, after confirmation that the entanglement entropy is infinite at the maximum, we study what happens at other critical points of the classical evolution (minima in oscillating models and the inflection points). We analyse the entanglement quantities for the universes with some exotic (non-Big-Bang) singularities in section \ref{EEExotic}, and in section \ref{EEHP} we suggest how the entanglement entropy behaves in the vicinity of singularities using our simple model. Finally, in section \ref{Conclusions}, we give our conclusions.

\section{Mathematical Formalism of third quantization}
\label{S2}

First of all, lets write the total action for gravity and a scalar field \cite{KieferBook}:
\begin{align}\label{0.1}
 S &=\frac12\int\d t N\left(-\frac{a\dt{a}^2}{N^2}+aK-\frac{\Lambda a^3}{3}\right)+ \nonumber \\
      &\quad +\frac12\int\d tNa^3\left(\frac{\dt{\phi}^2}{N^2}-2V(\phi)\right),
\end{align}
where $a$ is the scale factor, $\phi$ a scalar field, $N$ the lapse function, $K$ the curvature index, $\Lambda$ the cosmological constant and $V(\phi)$ the potential of the scalar field. From here we see that the canonical momenta are
\begin{equation}\label{0.2}
p_a=-\frac{a\dt{a}}{N},\qquad\qquad  p_{\phi}=\frac{a^3\dt{\phi}}{N},
\end{equation}
which let us write the associated Hamiltonian as
\begin{equation}\label{0.3}
H=\frac{N}{2}\left(-\frac{p_a^2}{a}+\frac{p_{\phi}^2}{a^3}-aK+\frac{\Lambda a^3}{3}+2a^3V(\phi)\right).
\end{equation}

Taking the quantization of the momenta like
\begin{equation}\label{0.4}
p_a^2:=-\frac{1}{a}\pdv{}{a}\left(a\pdv{}{a}\right),\qquad \qquad p_{\phi}^2:=-\pdv[2]{}{\phi},
\end{equation}
where $\hbar=1$, the most general Hamiltonian constraint we are going to consider, using the parametrization $\alpha=\ln(a)$, is written as
\begin{equation}\label{0.5}
\frac{\partial^2}{\partial\alpha^2}-\frac{\partial^2}{\partial\phi^2}
-\left[\e^{4\alpha}k
-\e^{6\alpha}\left(\frac{\Lambda}{3}+2V(\phi)\right)\right]=0.
\end{equation}
A very interesting case is the case for which the scalar field has no potential $V(\phi)$, which leads to a very simple Hamiltonian constraint, where the function into brackets in (\ref{0.5}) is only a function of the scale factor.

Lewis and Riesenfeld in Ref. \cite{LewisRiesenfeld} described how to find an invariant operator for the time-dependent quantum harmonic oscillator such that the number operator is constant \cite{LewisRiesenfeld}. Years after, Kim  \cite{KimLewis} found how to simplify the method by knowing any two linearly independent solutions on the diagonal representation. Our first task is to find the solutions to the Wheeler-DeWitt equation which for a minisuperspace $(a, \phi)$ built on the massless scalar field $\phi$ and the scale factor $a=\e^{\alpha}$ has the hyperbolic form
\begin{equation}
\label{1}
\left[\pdv[2]{}{\alpha}-\pdv[2]{}{\phi}+f(\alpha,\Lambda,...)\right]\Psi(\alpha,\phi)=0,
\end{equation}
which is analogous to the quantum harmonic oscillator equation, where $f(\alpha, \Lambda,...)$ is the function which depends on the scale factor and other parameters of the chosen model but the scalar field, since we consider $V(\phi)=0$, such as the cosmological constant $\Lambda$, and $\Psi(\alpha,\phi)$ is the wave function of the universe.

We opt for making the choice of $\Psi(\alpha,\phi)$ to be separable by Fourier modes like
\begin{equation}\label{2}
\Psi(\alpha,\phi)=\int_k\d k A(k)\e^{\pm\i k\phi}\varphi_{k}(\alpha),
\end{equation}
where $k$ refers to the number of a mode of the scalar field $\phi$, $A(k)$ is a weight function representing the distribution of the modes $k$, and $\varphi_{k}(\alpha)$ is the wave function of the universe for each mode $k$ that after inserting (\ref{2}) into (\ref{1}) fulfils the differential equation
\begin{equation}\label{3}
\left[\pdv[2]{}{\alpha}+k^2+f(\alpha,\Lambda,...)\right]\varphi_{k}(\alpha)=0.
\end{equation}
Taking $k$ as constant, the solutions of this equation are usually a combination of Bessel functions.

If we follow the analytic Frobenius method to find the general solution to (\ref{3}), we check that the solutions always exist, and that they strictly depend on the structure $\i k$, and hence the two linearly independent solutions $\varphi_{k}^{(1,2)}(\alpha)$ of (\ref{3}) can be found to be complex conjugated. Furthermore, since the solutions depend on the structure $\i k$, the transformation $k\to -k$ is equivalent to a complex conjugation $\i\to-\i$, and therefore the wave functions of both universes can be interpreted as a particle and an antiparticle due to the complex conjugation or because they share an opposite distribution of modes.

This is an indication that the modes of the scalar field can be distributed, as a very physical example, such that the positive modes are taken by one of the universes and the negative modes for the other, make us to take the global solutions in (\ref{2}) like
\begin{align}\label{4}
  \Psi^{(1)}(\alpha,\phi) & :=\int_0^{\infty}\d k A(k)\e^{-\i k\phi}\varphi_k^{(1)}(\alpha), \\
  \Psi^{(2)}(\alpha,\phi) & :=\int_{-\infty}^0\d k A(k)\e^{-\i k\phi}\varphi^{(2)}_{k}(\alpha),
\end{align}
where we use the symmetry of the solutions and we chose the sign for $\e^{\pm\i k\phi}$ as a plane wave that travels towards positive modes. Besides, we can prove that $\Psi^{(1,2)}(\alpha,\phi)$ are complex conjugated if and only if $A(k)$ is symmetric around $k=0$, which clearly shows the conservation of the energy of the whole system. Thus, $A(k)$ is going to be taken as a symmetric Gaussian distribution around $k=0$ with $\sigma$ its standard deviation -- taken as constant
\begin{equation}\label{5}
A(k)=\frac{1}{\sigma\sqrt{2\pi}}\e^{-\frac{k^2}{2\sigma^2}},
\end{equation}
which is more sophisticated than the choice made in Ref. \cite{PacketsKiefer}. The reconstruction of $\Psi(\alpha,\phi)$ from $\varphi_{k}(\alpha)$ cannot in general be made analytically, so we cannot get analytic results for even the simplest model we analyse, and the results also depend on $\sigma$. In view of that we perform just numerical calculations. Anyway, this method works generally in order to find solutions and afterwards the entanglement entropy of any model with the Hamiltonian constraint like (\ref{1}).

The algorithm starts with writing the Wheeler-DeWitt equation (\ref{1}) as a quantum harmonic oscillator equation
\begin{equation}
\label{10}
\ddt{\Psi}(\alpha,\phi)+\omega^2(\alpha,\phi)\Psi(\alpha,\phi)=0,
\end{equation}
where $\dt{\Psi}:=\partial_{\alpha}\Psi$, and we recognize the momentum $P_{\Psi}:=\dt{\Psi}$, and the frequency as
\begin{equation}
\label{11}
\omega^2(\alpha,\phi):=-\partial_{\phi}^2+f(\alpha,\Lambda,...).
\end{equation}
As soon as we are able to split the Hamiltonian constraint in two equations: the first for the scale factor
\begin{equation}
\label{6}
\left[\pdv[2]{}{\alpha}+f(\alpha,\Lambda,...)\right]\Psi(\alpha,\phi)=E_{\alpha}\Psi(\alpha,\phi),
\end{equation}
and the second for the scalar field
\begin{equation}
\label{7}
-\pdv[2]{}{\phi}\Psi(\alpha,\phi)=E_{\phi}\Psi(\alpha,\phi),
\end{equation}
recognizing $E_{\alpha}$ and $E_{\phi}$ as the energies associated for each variable, fulfilling $E_{\alpha}+E_{\phi}=0$, then the Wheeler-DeWitt equation is only dependent on the scale factor and the energy $E_{\phi}$ which is fixed. Hence, the frequency in (\ref{11}) is given by
\begin{equation}
\label{12}
\omega^2(\alpha,\phi)\equiv\omega^2(\alpha):= E_{\phi}+f(\alpha,\Lambda,...).
\end{equation}

In the following, we are assuming that the universes are created in pairs. The notation $\ket{U_-U_+}\equiv\ket{U_-}\ket{U_+}$ represents a state in which $\ket{U_-}$ has been typically interpreted as describing an expanding universe and $\ket{U_+}$ a contracting one. However, in terms of the physical time experienced by their internal observers, whose physical time variables are reversely related \cite{RP2019c, RP2020a}, they can alternatively be interpreted as two expanding universes with conjugated charge and parity properties of their matter contents. Thus, the state $\ket{U_-U_+}$ can equivalently be interpreted as representing a universe-antiuniverse pair. In the present development we shall assume that interpretation. At the beginning, the pair is in the ground state $\ket{00}$ for all representations. However, such a state does not represent here the 'no-particle' state, but the state of minimal excitation of the universes. In the invariant representation of the Hamiltonian that leads to \eqref{10} (see Appendix \ref{app01}), the state $|00\rangle$ remains being the ground state along the entire evolution of the universes. In other words, the universe-antiuniverse pair stays in the ground state of the invariant representation along the entire evolution of the universes. That is the boundary condition which we impose on the state of the universe-antiuniverse pair, which seems to be quite natural provided that the universe-antiuniverse pair does not interact with other universes of the multiverse. However, as the universes expands, i.e. as the scale factor grows, the ground state $|00\rangle$ splits into a linear combination of excited states of the instantaneous diagonal representation, which we shall assume that represents the state of the single universes at a given moment of time (i.e. at a given value of the scale factor).

Following Ref. \cite{Interuniversal} (cf. Appendix \ref{app01}) we can write that the invariant representation is described by the annihilation and the creation operators (or perhaps more appropriately the ''ladder operators'' $c_\pm$ and $c_\pm^\dag$)
\begin{eqnarray}
\label{15.0}
c_+&=&\sqrt{\frac12}\left[\frac{1}{R}\Psi(\alpha,\phi)+\i\left(R\dt{\Psi}-\dt{R}\Psi(\alpha,\phi)\right)\right], \\
\label{15.5}
c_-^{\dagger}&=&\sqrt{\frac12}\left[\frac{1}{R}\Psi(\alpha,\phi)-\i\left(R\dt{\Psi}-\dt{R}\Psi(\alpha,\phi)\right)\right],
\end{eqnarray}
where, in this specific case (see Ref. \cite{KimLewis} for further explanations), we can take
\begin{equation}\label{16}
R:=\sqrt{\Psi^{2}_{(1)}(\alpha,\phi)+\Psi^2_{(2)}(\alpha,\phi)},
\end{equation}
with $\Psi_{(1,2)}(\alpha,\phi)$ being any two real solutions of (\ref{1}). Since the functions in (\ref{4}) are complex conjugated satisfying the Wheeler-DeWitt equation (\ref{1}), the real and the imaginary parts of any of them are also solutions, such that both are real functions and a very nice candidates for the functions in (\ref{16}).

On the other hand, the diagonal representation of the Hamiltonian is given by
\begin{eqnarray}
\label{13.0}
b_{+}(\alpha)&:=&\sqrt{\frac{\omega(\alpha)}{2}}\left[\Psi(\alpha,\phi)+\frac{\i}{\omega(\alpha)}\dt{\Psi}(\alpha,\phi)\right],  \\
\label{13.5}
b_{-}^{\dagger}(\alpha)&:=&\sqrt{\frac{\omega(\alpha)}{2}}\left[\Psi(\alpha,\phi)-\frac{\i}{\omega(\alpha)}\dt{\Psi}(\alpha,\phi)\right].
\end{eqnarray}

The Bogoliubov transformation between representations (\ref{15.0})-(\ref{15.5})  and (\ref{13.0})-(\ref{13.5}) is \cite{Birrell:1982ix}
\begin{eqnarray}
\label{B17.0}
\hat{c}_-&=&\alpha_B\hat{b}_--\beta_B\hat{b}^{\dagger}_+, \\
\label{B17.5}
\hat{c}_-^{\dagger}&=&\alpha_B^{\ast}\hat{b}_-^{\dagger}-\beta_B^{\ast}\hat{b}_+,
\end{eqnarray}
where $\alpha_B$ and $\beta_B$ are the Bogoliubov coefficients, that written in terms of $R$ from (\ref{16}) and the frequency from (\ref{7}) are
\begin{eqnarray}
\label{17.0}
\alpha_B&=&\frac12\left[\frac{1}{R\sqrt{\omega}}+R\sqrt{\omega}-\frac{\i\dt{R}}{\sqrt{\omega}}\right], \\
\label{17.5}
\beta_B&=&-\frac12\left[\frac{1}{R\sqrt{\omega}}-R\sqrt{\omega}-\frac{\i\dt{R}}{\sqrt{\omega}}\right].
\end{eqnarray}

Therefore, the ground state of the invariant representation can be expanded in terms of vectors in the Fock space of the diagonal representation as
\begin{equation}\label{18}
\ket{00}_i=\frac{1}{\abs{\alpha_B}}\sum_{n=0}^{\infty}\left(\frac{\abs{\beta_B}}{\abs{\alpha_B}}\right)^n\ket{n_-n_+}_d,
\end{equation}
where the states $\ket{n_-}_d$ and $\ket{n_+}_d$ refer to the excited levels of the universe and the antiuniverse, respectively, in the diagonal representation. The diagonal representation is denoted by the index $d$, and the invariant representation by the index $i$.

The quantum state of each single universe of the universe-antiuniverse pair is described by the reduced density matrix that is obtained by tracing out from the composite quantum state the degrees of freedom of the partner universe. This yields
\begin{align}\label{19}
 \rho_- & =\sum_{n=0}^{\infty}\expval{\rho}{n_+}_d\propto \nonumber\\
  & \propto \sum_n\e^{\frac{\omega(\alpha)}{T(\alpha)}\left(n+\frac12\right)}\ket{n_-}\bra{n_-}_d,
\end{align}
where $\rho=\ket{0_-0_+}\bra{0_-0_+}$ in any representation, and $T(\alpha)$ is the temperature of entanglement of the obtained thermal state
\begin{equation}\label{19T}
T(\alpha)=\frac{\omega(\alpha)}{2\ln[\coth(r)]},
\end{equation}
where (cf. \cite{Interuniversal})
\begin{equation}
\label{20}
q:=\tanh(r)=\frac{\abs{\beta_B}}{\abs{\alpha_B}},
\end{equation}
whose values are into the interval $[0,1]$. For some fixed finite $R$, we can see that $q=1$, if $\omega=0$.

The state \eqref{19} shows that in the diagonal representation the quantum state of each single universe of the universe-antiuniverse pair is given by a non-separable or mixed state whose associated entropy of entanglement measures the degree of quantum correlations between the two universes. In Ref. \cite{Interuniversal} it was stated that the temperature of entanglement is a good measure of the quantumness of the state. Then, we shall use the temperature of entanglement \eqref{19T} as well as the entropy of entanglement associated to the state \eqref{19} as the measures of the quantumness of the state of the universes.

The entanglement entropy is then calculated as the von Neumann entropy of the system
\begin{align}\label{21}
 S_{\text{ent}}(\alpha)&  :=-\Tr[\rho\ln(\rho)]=\cosh^2(r)\ln\left[\cosh^2(r)\right]- \nonumber\\
 & \qquad  -\sinh^2(r)\ln\left[\sinh^2(r)\right].
\end{align}

Now, we need to fix the model and calculate the entropy of entanglement. If we want to calculate it for one dimensional scenario, we just need to set $E_{\phi}=0$ in (\ref{12}), and solve the Wheeler-DeWitt equation. The outcome will only be dependent on the frequency $\omega^2(\alpha,\phi)$ so the solutions which are also strongly dependent on the frequency.


We can notice that the entanglement entropy diverges when the frequency $\omega^2(a)$ is zero, since it plays the role of the potential in the Hamiltonian constraint (\ref{10}), and therefore $r$ in (\ref{21}) and the entropy of entanglement in (\ref{20}) are directly affected, giving $r=0$, and hence $S_{\text{ent}}\to\infty$. Something similar happens with the parameter $q$ in (\ref{21}), which approaches the unity when $\omega\to0$.

Furthermore, as $\omega^2(\alpha,\phi)$ is real, being negative for a classically permitted region and positive for the classically forbidden region, then $\omega(\alpha,\phi)$ can be real or purely imaginary, respectively, and from (\ref{21}) one can obtain that
\begin{widetext}
\begin{equation}\label{21X}
\cosh^2(r)=\frac{1}{1-q^2}
=\begin{cases}
  \frac{\left(\frac{1}{R}+R\abs{\omega}\right)^2+\dt{R}^2}{4\abs{\omega}}, & \mbox{if } \omega^2\ge0 \qquad \text{ (Classically Allowed)}\\
  -\frac{\frac{1}{R^2}+\left(R\abs{\omega}-\dt{R}\right)^2}{4\abs{\omega}R\dt{R}}, & \mbox{if } \omega^2<0  \qquad \text{ (Classically Forbidden)}
\end{cases},
\end{equation}
\end{widetext}
whence, after substitution into (\ref{20}), we check that the {\it entropy is real for the classically allowed region}, and imaginary for the classically forbidden one. This reveals a relation between the entanglement entropy and the classical availability of the universe for certain regions in the phase space $(\alpha,\phi)$. Unless $\omega^2(\alpha,\phi)$ reaches an extremum, 
a point where the entropy is infinite indicates a wall between the classical region and the quantum region, where the entropy of entanglement can even take complex values. 

If the universe shows a maximum where the entanglement entropy diverges, and at the initial singularity we assume that the universe is extremely quantum, it is clear that in the middle of its evolution there is, at least, a minimum of this entropy which seems, a priori, unrelated to any specific event of the evolution, though still it can have some meaning in the context of information theory which we have not yet discovered. 

Another interesting variable is the temperature defined in (\ref{19T}). It was stated in \cite{Interuniversal} that the temperature could be a better measurement of the quantumness of a system than the entanglement entropy. Lets analyse this temperature together with the entanglement entropy. Using equations (\ref{17.0}), (\ref{17.5}), (\ref{19T}) and (\ref{21}), the temperature  when the universe approaches a singularity ($\omega\to0$) can be approximated to
\begin{equation}\label{ApT}
\lim_{\omega\to0}T(\alpha)\approx\frac{1+R^2\dt{R}^2}{4R^2},
\end{equation}
since $R>0$ and $\omega^2>0$ for the classically allowed region. This temperature is never divergent around singularities if $R$ is finite or $R$ and $\dt{R}$ are not divergent at the same time, which is in general the case. Therefore, the temperature is finite in general.

In cases when the universe does not recollapse, by assumption the wave function of the universe is made to vanish at infinity, the parameter $R$ in (\ref{16}) also vanishes. Then the temperature, in general, diverges for most of the frequencies $\omega(\alpha)$.


Finally, it is worth exposing another alternatives to the entanglement entropy calculated as von Neumann entropy. First, we could track the behaviour of the parameter $q$ from (\ref{21}), and other entropic measurements like for example, Tsallis entropy $S^{\text{(T)}}_q$ \cite{Tsallis,TSALLIS2009} or Renyi entropy $S^{\text{(R)}}_q$ \cite{zbMATH03143975,Renyi}, which are defined and written in terms of $r$, as we did for the von Neumann entropy (\ref{21}), like
\begin{align}\label{21XX}
S_q^{\text{(T)}}(\rho)&:=\frac{1}{1-q}\left[\Tr(\rho^q)-1\right]=\nonumber\\
&=\frac{1}{1-q}\left[\frac{\sech^q(r)}{1-\tanh^{2q}(r)}-1\right],
\end{align}
and
\begin{align}\label{21XXX}
S^{\text{(R)}}_q(\rho)&:=\frac{1}{1-q}\ln[\Tr(\rho^q)]=\nonumber\\
&=\frac{1}{1-q}\ln\left[\frac{\sech^q(r)}{1-\tanh^{2q}(r)}\right],
\end{align}
respectively. These entropies are related to each other by 
\begin{equation}
\label{RenyiTsallis}
S^{\text{(R)}}_q =\frac{1}{1-q}\ln[1 + (1-q) S_q^{\text{(T)}}] .
\end{equation}


\section{Extended Minisuperspace Entanglement Entropy}\label{E2dim}

In order to extend the discussion of Ref. \cite{Interuniversal} into the matter degrees of freedom, we consider massless scalar field as another coordinate in minisuperspace as it was presented in Section \ref{S2}.

We consider a closed universe ($K=+1$) without the cosmological constant and a massless scalar field (i.e. no potential term, so $V(\phi)=0$) with an arbitrary distribution of modes $k$. In this scenario, we write (\ref{3}) as (see e.g. \cite{KieferBook})
\begin{equation}\label{8}
\left[\pdv[2]{}{\alpha}+k^2-\e^{4\alpha}\right]
\varphi_{k}(\alpha)=0,
\end{equation}
whose solutions are modified Bessel functions of the first kind with complex order
\begin{equation}\label{9}
\varphi_k^{(1,2)}(\alpha)\propto I_{\pm\i k}\left(\frac{\e^{2\alpha}}{2}\right),
\end{equation}
that are complex conjugated. To reconstruct the wave function of the universe as the first solution of (\ref{4}), we perform a numerical procedure including the distribution function (\ref{5}). The solutions we get for the wave function of the universe, both are complex conjugated as we explained in Section \ref{S2} if and only if, $A(k)$ is symmetric around $k=0$. Taking the distribution of the modes as in (\ref{5}), that condition is fulfilled. Once they are complex conjugated functions, the real and the imaginary part solve the differential equation (\ref{1}) by themselves, so we will use them as the input for the calculation of the invariant representation.

\begin{figure}[h]
  \includegraphics[width=0.45\textwidth]{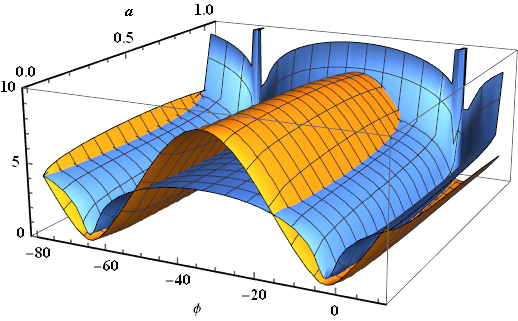}
  \caption{Entanglement Entropy (blue surface) and the temperature of entanglement (orange surface), of a pair of universes in a two dimensional minisuperspace ($a,\phi$). Here: $\sigma=1$. $E_{\phi}=1$. The entanglement entropy is infinite exactly at the maximum expansion diverging for any $\phi$, and finite for any point in the phase space of the classical evolution of the universe, even for $a=0$ at least for the precision of the numerical method we use. The temperature has been multiplied by a factor of $1/5$ to equalize both surfaces and it shows no divergences at any point of the phase space.}
  \label{Fig1}
\end{figure}

Using this numerical wave function, we calculate the parameter $R$ in (\ref{16}) and its derivative with respect to $\alpha$, the Bogoliubov coefficients using (\ref{17.0})-(\ref{17.5}), and the entropy of entanglement through the parameter $r$ from (\ref{20}) and (\ref{21}). The outcome, which is now given over a two-dimensional minisuperspace ($a,\phi$) is plotted in Fig. \ref{Fig1}.

Here we find a very important result. The entanglement entropy, which is a measurement of the quantumness of the system, is infinite at the maximum of expansion as we expected. However, the entropy at the Big-Bang singularity $a=0$ is not infinite. As a conclusion, we claim that according to our 2-dimensional minisuperspace calculations (the scale factor and the scalar field) the quantumness of the system is larger at maxima of the evolution of the universe than at the initial singularity, where $\dt{a}$ is non-vanishing, even though, it is important to notice that all entanglement entropy is locally decreasing in the neighborhood of the initial singularity. However, still in all the considered classical points (Big-Bang singularity, maxima), both the entropy and temperature of entanglement have non-zero values showing the entanglement of the universes and so some degree of quantumness of the system. Another speculation could be that more degrees of freedom in minisuperspace reduce the entanglement on a similar basis as it happens with the chaotic systems for which adding more particles can stabilise its behaviour and reduce disorder.


Furthermore, comparing with previous results for the entanglement entropy \cite{Interuniversal}, the main difference is the existence of a quantized scalar field, it is, we did not treat the scalar field classically. It seems now that it may regularise the value of the entropy of entanglement at the origin. However, it does not change the general interpretation. The generalised second principle of quantum thermodynamics \cite{Plenio1998} states that the entropy of entanglement cannot be increased by means of any local operation and classical communications alone. In the cosmological scenario, that means that the classical processes that happen inside a single universe, which in this context are all local irrespective of the distances between them\footnote{In the present context, a non-local operation is any that involve the composite state of the two universes of the entangled pair.}, cannot increase the entropy of entanglement between the pair of universes. Even more, they are expected to dissipate the quantum correlations between the states of the universes. Thus, the decreasing value of the entropy of entanglement from the very onset of the universes, regardless of whether the entropy of entanglement is infinite or not at the origin, would be associated to a classical evolution of the universes, which follows all the way up to the turning point of maximum expansion, where the universe enters again in a highly non-classical state \cite{PacketsKiefer, Interuniversal}. 

We have checked that the other measurements of entropy as Tsallis or Renyi entropy in (\ref{21XX}) and (\ref{21XXX}), respectively, are essentially analogous to the von Neumann entropy shown in Fig. \ref{Fig1}, and the same happens for the parameter $q$ from (\ref{21}) but the value at the divergences is unity. 


\section{Entanglement Entropy at Critical Points of Classical Evolution}
\label{EECP}

After our first result where the entanglement entropy diverges at the maximum point of the expansion of the universe, it is legitimate to ask if it always happens at any critical point of the oscillating evolution: the maxima, the minima or the inflection points.

\begin{figure}[h]
\centering
\includegraphics[width=0.85\linewidth]{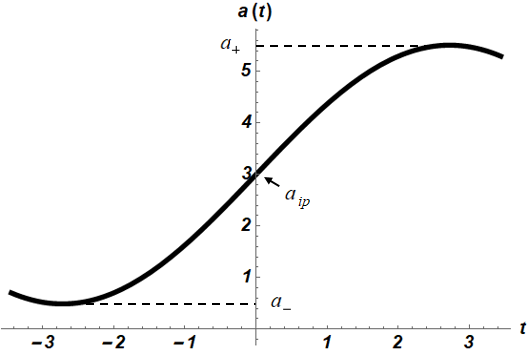}
\caption{Oscillation of the universe. Here: $\Lambda=-1$, $K=+1$, $C_s=0.1$, and $C_w=2$. The critical points are shown explicitly.\label{Fig2}}
\end{figure}

In order to figure out how the entropy behaves at these points, we apply a simple one-dimensional model such that the evolution of the universe shows all kind of the points: an oscillatory universe. A simple model explored in Ref. \cite{DABROWSKI1996} is the one with wall-like matter, string-like matter and a negative cosmological constant $\Lambda<0$. This model without string-like-matter was later reconsidered and dubbed a Simple Harmonic Universe (SHU) in the quantum context in Refs. \cite{Graham:2011nb,Mithani:2011en,Graham:2014pca,Mithani:2014jva,Mithani:2014toa} based on early Ref. \cite{Dabrowski:1995jt}. Some new aspects of stability of such models were recently studied in Ref. \cite{PhysRevD.100.083525}. In this model the scale factor behaves like
\begin{equation}
\label{22}
a(t)=-\frac{3}{2\Lambda}\left[A\sin\left(\sqrt{-\frac{\Lambda}{3}}t\right)+C_{w}\right], \end{equation}
where $C_w$ is a constant density parameter due to wall-like matter, such that
\begin{equation}\label{23}
C_w>A:=\sqrt{C_w^2+\frac43\Lambda k'},
\end{equation}
with $k'=K-C_s$, with $K$ being the curvature index and $C_s$ a constant density parameter due to string-like matter. The universe is oscillating between the minimum $(-)$ and the maximum $(+)$ defined by
\begin{equation}
\label{23.5}
a_{\pm}=-\frac{3}{2\Lambda}(\pm A+C_w).
\end{equation}
An example of the oscillation in this model is plotted in Fig. \ref{Fig2}.

\begin{figure}[h]
  \centering
  \includegraphics[width=0.45\textwidth]{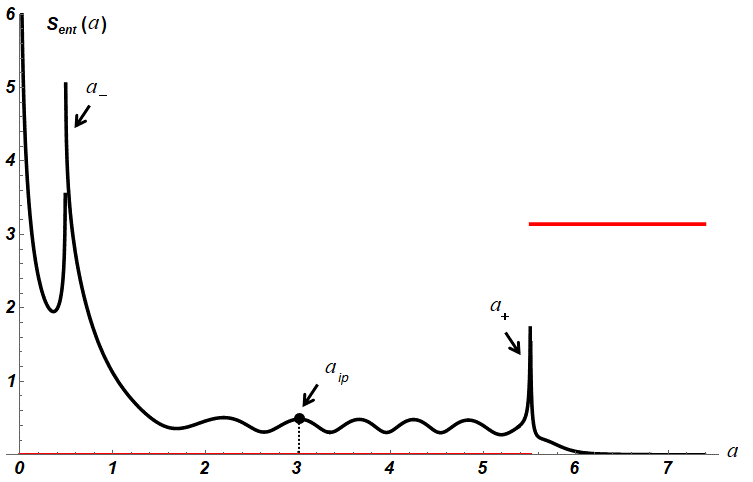}
  \caption{Entanglement entropy of an oscillating closed universe (\ref{22}). Here $a_{\pm}$ stand for the maximum and the minimum size of the oscillation as given by (\ref{23.5}), and $a_{\text{i.p.}}$ for the inflection point. We have chosen $\Lambda=-1$, $C_s=0.1$, and $C_w=2$. Black line represents the real part of the entropy while the red line its imaginary part. The entropy is finite though non-zero at the inflection point and diverges at those critical points where $\dot{a}=0$.}\label{Fig3}
\end{figure}

The Friedmann equation for this model is simply
\begin{equation}\label{24}
H^2=\frac{\Lambda}{3}+\frac{C_w}{a}-\frac{k'}{a^2}, \end{equation}
whence the Lagrangian is calculated as
\begin{equation}\label{25}
L=\frac12\int\d N a^3\left[\frac{H^2}{N^2}-\frac{\Lambda}{3}-\frac{C_w}{a}+\frac{k'}{a^2}\right],
\end{equation}
where $N$ is the lapse function. The corresponding Hamiltonian is
\begin{equation}
\label{26}
H=-\frac{p_a^2}{a}-k'a+C_wa^2+\frac{\Lambda}{3}a^3,
\end{equation}
and the Wheeler-DeWitt equation reads as
\begin{equation}
\label{27}
\left[-\hat{p}_a^2-k'a^2+C_wa^3+\frac{\Lambda}{3}a^4\right]\Psi(a)=0,
\end{equation}
where $\Psi(a)$ is the wave function of the universe. Quantizing the Hamiltonian with the proper factor ordering as described in (\ref{0.4}), one rewrites (\ref{27}) as
\begin{equation}
\label{29}
\left[\frac{1}{a}\dv{}{a}\left(a\dv{}{a}\right)-k'a^2+C_wa^3+\frac{\Lambda}{3}a^4\right]\Psi(a)=0.
\end{equation}
Using the parametrization $\alpha=\ln(a)$, it yields a kind of harmonic oscillator equation
\begin{equation}
\label{30}
\left[\dv[2]{}{\alpha}+\omega^2(\alpha)\right]\Psi(\alpha)=0, \end{equation}
with the frequency
\begin{equation}\label{31}
\omega^2(\alpha)=-k'\e^{4\alpha}+C_w\e^{5\alpha}+\frac{\Lambda}{3}\e^{6\alpha}.
\end{equation}
A priori, this equation has no easy analytic solution, but following the same procedure as in section \ref{S2} with $E_{\phi}=0$ and the solutions of (\ref{30}), we get the entropy of entanglement as plotted in Fig. \ref{Fig3}. As expected, we can see that the blow-up of the entanglement entropy coincides precisely with the minimum and the maximum of the expansion of the universe. However, the inflection point is shown to be of less importance since the entropy remains smooth (finite) around it. This result is a bit disappointing since the universe we inhabit has passed an inflection point between the matter-dominated and the dark energy-dominated era not so long ago.

\begin{figure}[h]
  \centering
  \includegraphics[width=0.45\textwidth]{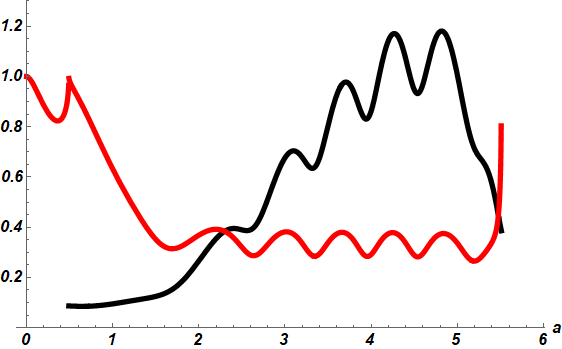}
  \caption{Representation of the temperature in black, and the parameter $q$ in red. The temperature has been multiplied by $1/10$ to make easier the comparison of the oscillations. The temperature is finite, showing weaker signal of the quantumness of the universe at any point.}\label{Fig3T}
\end{figure}

For completeness, we calculate the value of the entanglement temperature (\ref{19T}) for this model, and the result is plotted in Fig. \ref{Fig3T} together with the parameter $q$ from (\ref{21}). Both have been plotted qualitatively since their values are not so important as their behaviour. Again the temperature shows no divergences, while the parameter $q$ which is closely related to the entanglement entropy, maintains similar shape and it approaches unity close to the singularities.

\section{Entanglement of Exotic Singularity Universes}
\label{EEExotic}

\subsection{Other than Big-Bang singularities}

As one can consult from (see Table \ref{classif}), the first example of a non-Big-Bang type (from now on type 0) of a singularity in cosmology which is compatible with observations is a Big-Rip (BR or type I) associated with phantom dark energy \cite{Caldwell:1999ew,Dabrowski:2003jm}. Other examples are: a sudden future singularity (SFS or type II) \cite{Barrow:2004xh} which include Big-Break \cite{Exotic2007}, finite scale factor singularities (FSF or type III) which include Big-Freeze,  big-separation singularities (BS or type IV) \cite{Nojiri:2005sx}, and $w$-singularities (type V) \cite{Dabrowski:2009kg}. There is also a version of a Big-Rip known as a Little-Rip \cite{Frampton:2011sp}.
It seems fascinating that some of these singularities are ``weaker'' and some are  ``stronger'' than a Big-Bang in the sense of leading to infinities of some specific type and not to the other type \cite{Dabrowski:2009pc,Dabrowski:2014fha,Dabrowski:2018ucy}.

In order to understand what we mean by ``weaker'' and ``stronger'' we have to refer to some mathematical tools to investigate the problem. 
According to the definition of Tipler \cite{TIPLER19778} a singularity is strong if an extended object represented by three linearly independent, vorticity-free geodesic deviation vectors at $p$ parallely transported along causal geodesic $l$ is crushed to zero volume at the singularity by infinite tidal forces. In mathematical terms it means that at least one component of the tensor $I_j^i (\tau) = \int_0^{\tau} d\tau' \int_0^{\tau'} d\tau'' |R^i_{~ajb}u^a u^b|$ ($R^i_{~ajb}$ - the Riemann tensor, $u^a$ - four-velocity vector, $a,b,i,j = 0,1,2,3$, $\tau$ - proper time) diverges on the approach to a singularity at $\tau = \tau_s$. On the other hand, according to the definition of Kr\'olak \cite{Krolak_1986}, a singularity is strong if the expansion of every future-directed congruence of null (timelike) geodesics emanating from the point $p$ and containing $l$ becomes negative somewhere on $l$ or, in mathematical terms, if at least one component of the tensor $I_j^i(\tau) = \int_0^{\tau} d\tau' |R^i_{~ajb}u^a u^b|$ diverges on the approach to a singularity at $\tau = \tau_s$. For the null geodesics one replaces the Riemann tensor with the Ricci tensor components.

\begin{table*}
\caption{Classification of basic singularities in Friedmann cosmology. Here $t_s$ is the time when a singularity appears, $w=p/\rho$ is the barotropic index, $T$ - Tipler's definition, $K$ - Kr\'olak definition. In this paper we mainly concentrate on types 0, I$_l$, II$_a$, III$_a$, and IV.}
\label{classif}
\begin{center}
\begin{tabular}{lcccccccccc}
\hline
\\
Type & Name & $t$ & a(t$_s$) & $\varrho(t_s)$ & p(t$_s$) & $\dot{p}(t_s)$  & w(t$_s$) & T & K\\
\hline
\\
0  & Big-Bang (BB) & $ 0 $ & $ 0$ & $\infty$ & $\infty$ &$\infty$& finite & strong & strong\\
I  & Big-Rip (BR) & $t_s $ &$\infty$ & $\infty$ & $\infty$ &$\infty$& finite & strong & strong\\
I$_l$  & Little-Rip (LR) & $\infty$ &$\infty$ & $\infty$ & $\infty$ & $\infty$ & finite & strong & strong\\
II  & Sudden Future (SFS) & $t_s$ & $a_s$ & $\varrho_s$ & $\infty$ & $\infty$ & finite & weak & weak\\
II$_a$& Big-Brake (BBr) & $t_s$ & $a_s$ & $0$ & $\infty$ & $\infty$ & finite & weak & weak \\
III  & Finite  Scale  Factor (FSF) & $t_s$ &$a_s$ & $\infty$ & $\infty$ &$\infty$& finite& weak & strong\\
III$_a$  & Big-Freeze (BF) & $t_s$ & $0$ & $\infty$ & $\infty$ &$\infty$& finite& weak & strong\\
IV & Big-Separation  (BS) & $t_s$ &$a_s$ & $0$ & $0$ &$\infty$& $\infty$ & weak & weak\\
V & w-singularity (w) & $t_s$ &$a_s$ & $0$ & $0$ &$0$& $\infty$& weak & weak
\\
\hline
\end{tabular}
\end{center}
\end{table*}




\subsection{Type II singularity universes entanglement}

Physically, it is a singularity for which the tidal forces manifest here as the (infinite) impulse which reverses for SFS (or stops to zero for Big-Brake) the increase of separation of geodesics and the geodesics themselves can evolve further -- the universe can continue its evolution through a singularity. This behaviour is like a turning point of a harmonic oscillator. In a specific example of type II singularity - Big-Brake - the effect of a scalar field which fulfils the equation of state of a generalized Chaplygin gas, that is
\begin{equation}
\label{E0}
p=-\frac{A}{\rho^{\beta}},
\end{equation}
where $p$ is the pressure, $\rho$ the density, $A<0$ a constant, and throughout this section $\beta=-1$.  The dependence of the scalar field with the scale factor is found through the continuity equation which takes the form
\begin{equation}\label{E1}
\rho=\sqrt{\frac{B}{a^6}-A},
\end{equation}
where $B>0$, and whence at the singularity time $t_s$ one has $a_s = a(t_s) =(B/A)^{1/6}$. The evolution of the scale factor begins with the Big-Bang singularity at time $t=0$ where $a(0) = 0$ and stops at the Big-Brake singularity at time $t_s$ where $a(t_s) = a_s >0$. The exotic Big-Brake singularity that appears in this model has the properties
\begin{equation}\label{E1.5}
\dt{a}(t_s)=0,\quad \ddt{a}(t_s)\to-\infty,\quad \frac{A}{p}=\rho(t_s)\to 0.
\end{equation}
The scalar field which leads to such a singularity is treated as classical, so we are going to proceed with quantization using  (\ref{0.3}) with $N=1$, i.e.
\begin{equation}\label{E2}
H=\frac{1}{2}\left(-\frac{p_a^2}{a}-aK+\frac{\Lambda a^3}{3}+2a^3\rho\right),
\end{equation}
and after quantization, the Wheeler-DeWitt equation is
\begin{equation}\label{E3}
\left[\pdv[2]{}{\alpha}-\e^{4\alpha}K+\e^{6\alpha}\left(\frac{\Lambda}{3}+2\rho(\alpha)\right)\right]\Psi(\alpha)=0,
\end{equation}
where, in the case described in \cite{Exotic2007}, there is no cosmological constant ($\Lambda=0$), the density of the scalar field $\rho(\alpha)$ is as in (\ref{E1}), and the flat case $(K=0)$ is considered.

\begin{figure}[h]
  \includegraphics[width=0.45\textwidth]{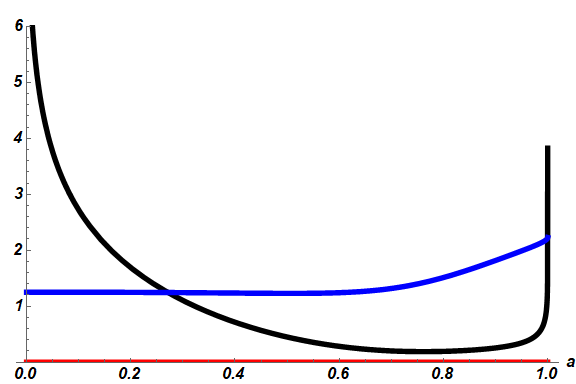}
  \caption{Entanglement entropy and temperature for a model which scalar field follows the equation of state (\ref{E0}). We have chosen: $B=-A=1$. There is a divergence at the Big-Brake singularity for $a=a_s=1$ as well as for the initial Big-Bang singularity where $a=0$. The black line is the real part of the entropy, and the red one is the imaginary part. The temperature (blue) has been multiplied by a factor of 5 to make the comparison easier.}
  \label{FigE1}
\end{figure}

The entanglement entropy for $B=-A=1$, is plotted in Fig. \ref{FigE1}. There we see the divergence at the initial Big-Bang singularity ($a=0$) as well as at Big-Brake singularity at $(a=a_s=1)$. Again the temperature does not show something relevant, but its value is non-zero.

\subsection{Type III singularity universes entanglement}\label{STypeIII}

Another exotic singularity - the Big-Freeze - is the one studied in \cite{Exotic2009}, which appears when there is a scalar field with polytropic equation of state (\ref{E0}), where $\beta$ is any constant. It is a special case of type III singularity. If the model is considered for a flat universe without cosmological constant, then the density for which the null, strong and weak energy conditions are fulfilled, can be written in terms of the scale factor as
\begin{equation}
\label{E5}
\rho=\abs{A}^{1/(1+\beta)}\left[\left(\frac{a_{s}}{a}\right)^{3(1+\beta)}-1\right]^{1/(1+\beta)},
\end{equation}
where
\begin{equation}\label{E6}
a_{s} = a(t_s) =\abs{\frac{B}{A}}^{1/3(1+\beta)},
\end{equation}
is the minimum size of the universe, $A<0$, $1+\beta<0$, and $B>0$ is a constant of integration. At $a_{s}$ is where the initial Big-Freeze singularity appears, and its properties are
\begin{equation}\label{E7}
\rho(a_{s}),p(a_{s})\to\infty, \qquad \dt{a}(a_{s})\to\infty.
\end{equation}
Besides, there is no maximum size, and the minimum size is not zero in general, and it occurs at a finite time $t_s$.

\begin{figure}[h]
  \includegraphics[width=0.45\textwidth]{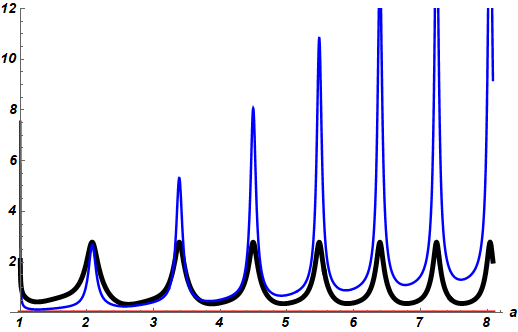}
  \caption{Entanglement entropy and temperature for a model which scalar field follows the equation of state (\ref{E5}). We have chosen: $A=-1$, $B=1$ and $\beta=-2$. There is a divergence for the Big-Freeze singularity, here at $a=a_s=1$. The black line is the real part of the entropy, and the red one the imaginary part. The blue line shows the temperature multiplied by $1/10$, which increases as the universe expands.}
  \label{FigE2}
\end{figure}

Again, the scalar field is treated classically, so we stick to the Wheeler-DeWitt equation (\ref{E3}) to proceed.  The entanglement entropy in Fig. \ref{FigE2} shows nothing special, even a divergence at $a_{s}$ that was already expected since $\rho(a_{s})\to\infty$, and therefore the frequency $\omega^2(\alpha)$ in (\ref{E3}) also diverges, the parameter $q$ in (\ref{21}) approaches unity, and finally the entanglement entropy in (\ref{20}) diverges. Since there is no maximum size for this universe, the envelope of the oscillations of the entanglement entropy decreases as the universe grows. 

As the universe expands, the temperature, also showed in Fig. \ref{FigE2}, increases since $R$ in (\ref{16}) goes to zero due to the vanishing behaviour of the wave function, which is imposed as a boundary condition for the wave function of the universe. A priori, it is said that the quantumness of the universe vanishes when the universe expands. Hence, the temperature cannot be a good indicator of the quantumness, at least in this scenario.

\subsection{Big-Separation (type IV) singularity universes entanglement}

The model for which the next exotic singularity arises also takes into account a flat universe without cosmological constant, and a scalar field which equation of state is (\ref{E0}), where $A<0$ and $\beta$ is any constant. The density in terms of the scale factor which fulfils all the energy conditions is (\ref{E5}), where $B>0$, and $\beta\in(-1/2,0)$,
but $a_{s}$ is now the maximum of the expansion of the universe defined by (\ref{E6}).

The type IV singularity appears at $a=a_{s}$ for $t=t_s$, and its characteristics are
\begin{equation}\label{E8}
p(t_{s}),\rho(t_s)=0, \qquad \dt{a}(t_s)=0.
\end{equation}

\begin{figure}[h]
  \includegraphics[width=0.45\textwidth]{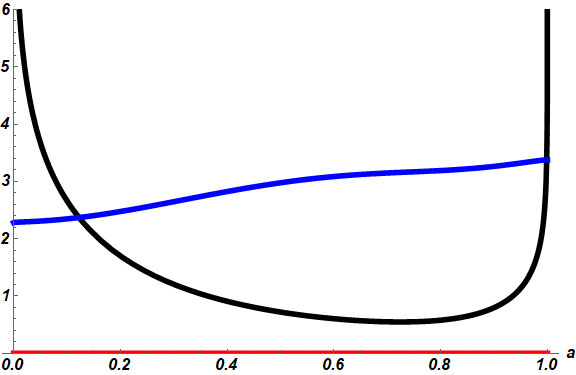}
  \caption{Entanglement entropy and temperature for a model which scalar field follows the equation of state (\ref{E0}). We have chosen: $A=-1$, $B=1$ and $\beta=-1/2$. There is a divergence in the type IV singularity, here at $a=a_s=1$. The black line is the real part of the entropy, and the red one the imaginary part. The blue line is the temperature, which has been multiplied by a factor of 10. It shows no special behaviour at any singularity.}
  \label{FigE3}
\end{figure}

The entanglement entropy is calculated from the Hamiltonian constraint (\ref{E3}) and represented in Fig. \ref{FigE3} together with the temperature, which shows a divergence at the initial singularity and another at the type IV singularity. It was expected since $\dt{a}(t_s)=0$, because the universe reaches a maximum at $a_{s}$. The temperature shows no divergences at any point, remaining approximately constant during the evolution. The parameter $q$ behaves as expected, it is the unity at $a=0$ and $a=a_s=1$, and it follows the same shape as the entanglement entropy.

\subsection{Little-Rip singularity entanglement}

\begin{figure}[h]
  \includegraphics[width=0.45\textwidth]{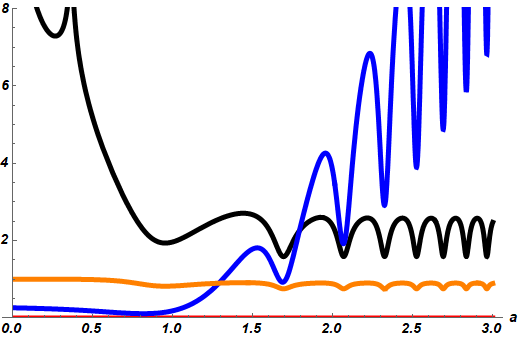}
  \caption{Entanglement entropy and temperature for a model which scalar field follows the equation of state (\ref{E9}). We have chosen: $A=2/3$, $B=1$, $\rho_o=1$, and $a_o=1$. There are two divergences: the initial singularity at $a=0$, and another at $a=a_{s}$. As the universe approaches the Little-Rip event at $a\to\infty$, nothing special happens. The black line is the real part of the entropy, and the red one the imaginary part. The temperature, in blue, increases and oscillates as the universe expands. It has been multiplied by a factor 1/20. The orange curve is the parameter $q$, which is around the unity in the beginning of the evolution, the unity at the divergences, and it is oscillating around the value $0.8$ as the universe expands.}
  \label{FigE4}
\end{figure}

The Little-Rip event described in \cite{Exotic2016} is an event for which the scale factor, the Hubble parameter $H$ and its derivative with respect to the cosmic time diverge, which is not precisely a singularity. This can be produced by a scalar field with equation of state like
\begin{equation}\label{E9}
p=-\rho-A\sqrt{\rho}<0,
\end{equation}
where $A>0$ is a constant. If the universe contains this scalar field, its curvature is flat and has no cosmological constant, the density can be written in terms of the scale factor as
\begin{equation}\label{E10}
\rho=\rho_o\left[\frac{3A}{2\sqrt{\rho_o}}\ln\left(\frac{a}{a_o}\right)+1\right]^2,
\end{equation}
where $\rho_o$ and $a_o$ are constants of integration which correspond to the value of the density and the scale factor nowadays, respectively. The density in (\ref{E10}) presents an interesting point at
\begin{equation}\label{E11}
a_{s}=a_o\e^{-\frac{2\sqrt{\rho_o}}{3A}},
\end{equation}
where the density is zero. Looking at the Wheeler-DeWitt equation (\ref{E3}), with $K=0$ and $\Lambda=0$, and comparing with (\ref{10}), we see that the frequency $\omega^2(\alpha)$ is directly proportional to the density (\ref{E10}), and therefore the frequency will also be zero, so a divergence of the entanglement entropy is expected there. Fig. \ref{FigE4} shows the temperature and the entanglement entropy of this model, for which this singularity appears due to the vanishing of the density of the scalar field. Besides, there is the initial singularity at $a=0$, but no divergence as the universe is getting closer to the little rip event, where the entanglement entropy is infinitely oscillating and globally decreasing.

The temperature and the parameter $q$ are also shown in Fig. \ref{FigE4}. The temperature is increasing for high values of the scale factor. This is again meaningful, because the universe is expected to lose its quantumness when it expands, and thus, the temperature should decrease if it was a good indicator of the quantumness of the universe, but that is not the case and the temperature increases as the universe evolves. The parameter $q$ is one at the singularities, and for the rest, it moves slightly oscillating around $0.8$.

\section{Entanglement Entropy Versus the Hubble Parameter}
\label{EEHP}

Since the entanglement entropy diverges at maxima and minima, the question is if it is related somehow with the Hubble parameter, since $H:=\partial_{a}\ln(a)$ is null at those points.  In order to analyse this problem, we take the simplest minisuperspace model ($a,\phi$), described by (\ref{8}), that is the universe with only a scalar field and a unique mode $k=1$. The wave functions for this model are those of (\ref{9}) with mode $k=1$.

\begin{figure}[h]
\centering
  \includegraphics[width=0.45\textwidth]{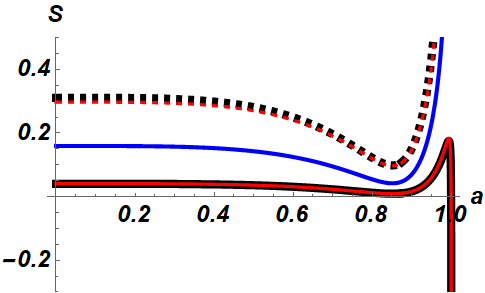}
  \caption{Different entropies vs. the scale factor. The blue line shows the von Neumann entropy, the solid lines show the Renyi (red) and Tsallis (black) entropies for $q=1.5$, and the dashed ones for $q=0.7$, respectively. For values $q\in(0,1)$, it diverges to minus infinity.}
  \label{FigEnts}
\end{figure}

The Lagrangian of the model is
\begin{equation}\label{32}
L=\frac12 \int \d N a^3 \left[\frac{H^2}{N^2}+\frac{1}{a^2}-\frac{\dt{\phi}^2}{N^2}\right],
\end{equation}
from where we get
\begin{equation}\label{33}
p_{\phi}=\dt{\phi}a=k\equiv \text{const},
\end{equation}
which is the mode of the scalar field. Therefore, the Friedmann equation is simply
\begin{equation}\label{34}
H^2-\frac{k^2}{a^6}+\frac{1}{a^2}=0,
\end{equation}
whence we obtain that, classically, the maximum size of the universe is at $a=1$ if $k=1$ (note that $k$ here means the number of the mode, while the curvature index $K=+1$ is ''hidden'' in the term $1/a^2$). The entanglement entropy is shown in Fig. \ref{FigEnts} with a blue line for this model. Again we found a divergence at the maximum of the evolution of the universe, and as in section \ref{E2dim}, a finite initial entanglement entropy.

\begin{figure}[h]
\centering
  \includegraphics[width=0.45\textwidth]{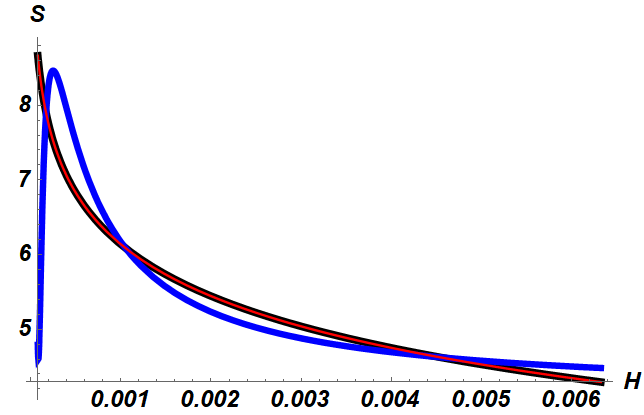}
  \caption{Entanglement entropy vs. the Hubble parameter. Black line is the exact entanglement entropy, while blue line is the fit for the entropy as (\ref{XXX}) and red line is the fit for the entropy given by the relation (\ref{XX}).}
  \label{Fig5}
\end{figure}

As our task is to look for the relation with the Hubble parameter, we use (\ref{34}) to rewrite the entropy of entanglement numerically and perform a fit as we wish. Our objective is to find a relation like
\begin{equation}\label{X}
S_{\text{ent}}(H)\sim\frac{1}{H^2},
\end{equation}
since it will indicate a relation to the entropy of horizons, such as the Hubble horizon or the black hole horizon. In order to check if it the case, we approximate the entropy very close to a singularity by the function
\begin{equation}
\label{XXX}
S_{\text{ent}}(H)\sim
  c_o+\frac{c_1}{H}+\frac{c_2}{H^2}+\frac{c_3}{H^3},
\end{equation}
where we expect the condition $c_2\gg c_o,c_1,c_3$ to be valid. The result of the fit is shown in Fig. \ref{Fig5} with a an unexpected result. The entropy of entanglement seems to fit perfectly to a different function
\begin{equation}\label{XX}
S_{\text{ent}}(H)\sim c_o-c_1\ln(H),
\end{equation}
where $c_1\approx 1$, and we can see that this entropy has the shape of the Shannon entropy \cite{Shannon} for an event with probability $H$. The fit can be found in Fig. \ref{Fig5}. 

\begin{figure}[h]
\centering
  \includegraphics[width=0.45\textwidth]{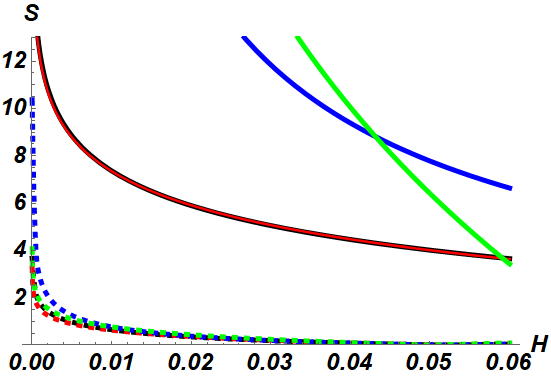}
  \caption{Different entropies vs. the Hubble parameter. The solid lines show the Renyi (black) and Tsallis (blue) entropies in terms of $H$ for $q=0.7$, and their fits in red and green, respectively. The dashed lines show absolute values of the Renyi (black) and Tsallis (blue) entropies for $q=1.5$, with their fits in red and green, respectively. }
  \label{FigFitEnts}
\end{figure}

We wanted to check if this logarithmic shape holds for alternatives of the von Neumann entropy, as Renyi or Tsallis entropies. Using (\ref{21XX}) and (\ref{21XXX}), we calculated both entropies and drew them in Fig. \ref{FigEnts}, together with von Neumann entropy. The blue line shows the von Neumann entropy, the solid lines show the Renyi (red) and Tsallis (black) entropies for $q=1.5$, and the dashed ones for $q=0.7$, respectively. The fits of the form (\ref{XX}) of these alternative entropies are in Fig. \ref{FigFitEnts}. The solid lines shown the Renyi (black) and Tsallis (blue) entropies in terms of $H$ for $q=0.7$, and their fits in red and green, respectively. The dashed lines show the absolute values of the Renyi (black) and Tsallis (blue) entropies for $q=1.5$, with their fits in red and green, respectively.

The fits of Renyi and Tsallis entropies to a function like (\ref{XXX}) or (\ref{XX}) are not as appropriate as for von Neumann entropy. Nevertheless, as shown in Fig. \ref{FigEnts}, all of them behave similarly if $q>1$, and therefore they could be taken as good indicators of the quantumness of the universe. They even yields an imaginary part out of the classical region as the von Neumann entropy. If $q\in(0,1)$, the entropy diverges to -$\infty$ at the maximum expansion of the universe, and a maximum of entropy appears, which is also unrelated to any special event.

\section{Conclusions}
\label{Conclusions}

We have used the 3rd quantization formalism to calculate the entanglement entropy of two universes created in pairs within the framework of the Friedmann cosmology. Our main concern has been to calculate the entanglement entropy and the entanglement temperature around maxima, minima and inflection points of the classical evolution. This has been done in an extended two-dimensional minisuperspace parameterised by the scale factor $a$ and the massless scalar field $\phi$.

We have found that after the Fourier expansion of the wave function of the universe $\Psi(\alpha,\phi)$ like (cf. (\ref{2})), the only way to have two complex conjugated functions $\Psi^{(1,2)}(\alpha,\phi)$ representing both the universe and the anti-universe, is to have symmetric distribution $A(k)$ of the modes of the scalar field around $k=0$. This is quite strong, though necessary, constraint on the distribution of the scalar field into a pair of universes. This has proven the symmetry of the system, i.e. that the energy is conserved.

We have found that, while taking a scalar field $\phi$ into consideration, the entanglement entropy is finite at the classical place of the Big-Bang singularity, and diverges at maxima and minima of expansion. We cannot give, a priori, any reason to the finite value of the entanglement entropy at the initial singularity, but it seems to be related to the inclusion of another degree of freedom such as the quantized scalar field. Indeed, when we consider the scalar field as classical and give some relation between its energy and the scale factor as we did in Sections \ref{EECP} and \ref{EEExotic}, the scalar field is no longer an independent variable, and the entanglement entropy blows up at that singular point.

It is worth mentioning that every statement about a critical point or a singularity which one infers from the canonical quantum cosmology is dependent on the construction of the theory far from the semiclassical point of view. The problem is the factor ordering which affects small scales and/or high curvatures, and makes the quantization as we took it in (\ref{0.4}) to be one of the choices one can make, but there are infinitely many other possibilities which could lead to completely different results (see eg. \cite{TeiglFactor}).

We have also studied the entanglement quantities for the universes which classically exhibit other than Big-Bang singularities such as Big-Brake, Big-Freeze, Big-Separation and Little-Rip. We have found that the entanglement entropy or the entanglement temperature in all the critical points and singularities is either finite or infinite but it never vanishes. This has proven that the universes within each pair are entangled and so subject to some quantumness within the system. Furthermore, we have found in Section \ref{EEExotic}, when we analysed the Big-Freeze singularity, that an earlier statement made in Ref. \cite{Interuniversal} about the temperature as a proper measure of the quantumness is not always true, since in this specific example, the temperature diverges when the universe expands infinitely., while it was expected that the entanglement of a pair of universes should vanish asymptotically as the universe expands. 

Apart from the von Neumann entanglement entropy, we have also calculated the Tsallis and Renyi entanglement entropies and found that they exhibit similar behaviour as the measures of the quantumness as the von Neumann entropy. This seems to show the robustness of our entanglement calculations and the appropriate conclusions. 

Finally, we have been looking for the relation between the entanglement entropy and the Hubble parameter which classically marks special points of the evolution of the universe. In the example analysed in Section \ref{EEHP}, we have found the logarithmic relation (\ref{XX}) to be a very good approximation to the behaviour of the entanglement entropy in terms of the Hubble parameter around critical points, where the Hubble parameter vanishes. Further investigations should be aimed to find explanations and consistency of this result.

\appendix
\section{Invariant Representation of Equation (\ref{10})}\label{app01}

The invariant representation of the Hamiltonian that in the third quantization formalism leads to the Wheeler-DeWitt equation \eqref{10} (with the frequency given by \eqref{12}) can easily be obtained in the following way. Let us firs notice that if we make the change in the wave function of the universe, $\Psi \rightarrow \xi = \frac{1}{R} \Psi$, with $R$ being a function that satisfies
\be
\ddot R + \omega^2 R = \frac{1}{R^3} ,
\ee
and the change in the scale factor, $\alpha \rightarrow \tilde \alpha$, defined by
\be
\tilde\alpha(\alpha) = \int^\alpha \frac{d \alpha'}{R^2(\alpha')} ,
\ee
then, the Wheeler-DeWitt equation \eqref{10} transforms into
\be\label{CHO}
\xi'' + \xi = 0 ,
\ee
where, $\xi' \equiv \frac{\partial\xi}{\partial \tilde \alpha}$. Eq. \eqref{CHO} is the equation of a harmonic oscillator with constant unit frequency. The associated creation and annihilation operators, defined as usual by
\beq\label{I01}
c_\pm &=& \sqrt{\frac{1}{2}} \left( \xi + i \xi' \right) , \\  \label{I02}
c_\pm^\dag &=& \sqrt{\frac{1}{2}} \left( \xi - i \xi' \right) ,
\eeq
determine a representation $| N\rangle$, given by the eigenstates of the number operator $N = c_\pm^\dag c_\pm$, that is invariant under the evolution of the scale factor. Now, taking into account that
\be
\xi = \frac{1}{R} \Psi  ,
\ee
and
\be
\xi' \equiv \frac{\partial \xi}{\partial \tilde \alpha}Â = R \dot \Psi - \dot R \Psi ,
\ee
one can write the operators (\ref{I01})-(\ref{I02}) of the invariant representation as (\ref{15.0})-(\ref{15.5}).

\section*{Acknowledgments}
S.B.B would like to thank F. Wagner for his decisive help and long conversations which gave rise to Section \ref{EEHP}, and A. Mart{\'i}n-Gal{\'a}n for her help and discussions about Section \ref{S2}.

The work of S.B.B. was supported by the Polish National Research and Development Center (NCBR) project ''UNIWERSYTET 2.0. --  STREFA KARIERY'', POWR.03.05.00-00-Z064/17-00 (2018-2022).

\newpage

\bibliography{multiv_critical_16.bib}

\begin{thebibliography}{52}%
\makeatletter
\providecommand \@ifxundefined [1]{%
 \@ifx{#1\undefined}
}%
\providecommand \@ifnum [1]{%
 \ifnum #1\expandafter \@firstoftwo
 \else \expandafter \@secondoftwo
 \fi
}%
\providecommand \@ifx [1]{%
 \ifx #1\expandafter \@firstoftwo
 \else \expandafter \@secondoftwo
 \fi
}%
\providecommand \natexlab [1]{#1}%
\providecommand \enquote  [1]{``#1''}%
\providecommand \bibnamefont  [1]{#1}%
\providecommand \bibfnamefont [1]{#1}%
\providecommand \citenamefont [1]{#1}%
\providecommand \href@noop [0]{\@secondoftwo}%
\providecommand \href [0]{\begingroup \@sanitize@url \@href}%
\providecommand \@href[1]{\@@startlink{#1}\@@href}%
\providecommand \@@href[1]{\endgroup#1\@@endlink}%
\providecommand \@sanitize@url [0]{\catcode `\\12\catcode `\$12\catcode
  `\&12\catcode `\#12\catcode `\^12\catcode `\_12\catcode `\%12\relax}%
\providecommand \@@startlink[1]{}%
\providecommand \@@endlink[0]{}%
\providecommand \url  [0]{\begingroup\@sanitize@url \@url }%
\providecommand \@url [1]{\endgroup\@href {#1}{\urlprefix }}%
\providecommand \urlprefix  [0]{URL }%
\providecommand \Eprint [0]{\href }%
\providecommand \doibase [0]{http://dx.doi.org/}%
\providecommand \selectlanguage [0]{\@gobble}%
\providecommand \bibinfo  [0]{\@secondoftwo}%
\providecommand \bibfield  [0]{\@secondoftwo}%
\providecommand \translation [1]{[#1]}%
\providecommand \BibitemOpen [0]{}%
\providecommand \bibitemStop [0]{}%
\providecommand \bibitemNoStop [0]{.\EOS\space}%
\providecommand \EOS [0]{\spacefactor3000\relax}%
\providecommand \BibitemShut  [1]{\csname bibitem#1\endcsname}%
\let\auto@bib@innerbib\@empty
\bibitem [{\citenamefont {DeWitt}(1967)}]{DeWitt1}%
  \BibitemOpen
  \bibfield  {author} {\bibinfo {author} {\bibfnamefont {B.~S.}\ \bibnamefont
  {DeWitt}},\ }\href {\doibase 10.1103/PhysRev.160.1113} {\bibfield  {journal}
  {\bibinfo  {journal} {Phys. Rev.}\ }\textbf {\bibinfo {volume} {160}},\
  \bibinfo {pages} {1113} (\bibinfo {year} {1967})}\BibitemShut {NoStop}%
\bibitem [{\citenamefont {Kiefer}(2012)}]{KieferBook}%
  \BibitemOpen
  \bibfield  {author} {\bibinfo {author} {\bibfnamefont {C.}~\bibnamefont
  {Kiefer}},\ }\href {\doibase 10.1093/acprof:oso/9780199585205.001.0001}
  {\emph {\bibinfo {title} {{Quantum gravity; 3rd ed.}}}},\ International
  series of monographs on physics\ (\bibinfo  {publisher} {Oxford Univ.
  Press},\ \bibinfo {address} {Oxford},\ \bibinfo {year} {2012})\BibitemShut
  {NoStop}%
\bibitem [{\citenamefont {Vilenkin}(1989)}]{PhysRevD.39.1116}%
  \BibitemOpen
  \bibfield  {author} {\bibinfo {author} {\bibfnamefont {A.}~\bibnamefont
  {Vilenkin}},\ }\href {\doibase 10.1103/PhysRevD.39.1116} {\bibfield
  {journal} {\bibinfo  {journal} {Phys. Rev. D}\ }\textbf {\bibinfo {volume}
  {39}},\ \bibinfo {pages} {1116} (\bibinfo {year} {1989})}\BibitemShut
  {NoStop}%
\bibitem [{\citenamefont {Hosoya}\ and\ \citenamefont {Morikawa}(1989)}]{3rdQ}%
  \BibitemOpen
  \bibfield  {author} {\bibinfo {author} {\bibfnamefont {A.}~\bibnamefont
  {Hosoya}}\ and\ \bibinfo {author} {\bibfnamefont {M.}~\bibnamefont
  {Morikawa}},\ }\href {\doibase 10.1103/PhysRevD.39.1123} {\bibfield
  {journal} {\bibinfo  {journal} {Phys. Rev. D}\ }\textbf {\bibinfo {volume}
  {39}},\ \bibinfo {pages} {1123} (\bibinfo {year} {1989})}\BibitemShut
  {NoStop}%
\bibitem [{\citenamefont {Holman}\ \emph
  {et~al.}(2008{\natexlab{a}})\citenamefont {Holman}, \citenamefont
  {Mersini-Houghton},\ and\ \citenamefont {Takahashi}}]{1}%
  \BibitemOpen
  \bibfield  {author} {\bibinfo {author} {\bibfnamefont {R.}~\bibnamefont
  {Holman}}, \bibinfo {author} {\bibfnamefont {L.}~\bibnamefont
  {Mersini-Houghton}}, \ and\ \bibinfo {author} {\bibfnamefont
  {T.}~\bibnamefont {Takahashi}},\ }\href {\doibase 10.1103/PhysRevD.77.063510}
  {\bibfield  {journal} {\bibinfo  {journal} {Phys. Rev. D}\ }\textbf {\bibinfo
  {volume} {77}},\ \bibinfo {pages} {063510} (\bibinfo {year}
  {2008}{\natexlab{a}})}\BibitemShut {NoStop}%
\bibitem [{\citenamefont {Holman}\ \emph
  {et~al.}(2008{\natexlab{b}})\citenamefont {Holman}, \citenamefont
  {Mersini-Houghton},\ and\ \citenamefont {Takahashi}}]{2}%
  \BibitemOpen
  \bibfield  {author} {\bibinfo {author} {\bibfnamefont {R.}~\bibnamefont
  {Holman}}, \bibinfo {author} {\bibfnamefont {L.}~\bibnamefont
  {Mersini-Houghton}}, \ and\ \bibinfo {author} {\bibfnamefont
  {T.}~\bibnamefont {Takahashi}},\ }\href {\doibase 10.1103/PhysRevD.77.063511}
  {\bibfield  {journal} {\bibinfo  {journal} {Phys. Rev. D}\ }\textbf {\bibinfo
  {volume} {77}},\ \bibinfo {pages} {063511} (\bibinfo {year}
  {2008}{\natexlab{b}})}\BibitemShut {NoStop}%
\bibitem [{\citenamefont {Valentino}\ and\ \citenamefont
  {Mersini-Houghton}(2018)}]{2.5}%
  \BibitemOpen
  \bibfield  {author} {\bibinfo {author} {\bibfnamefont {E.~D.}\ \bibnamefont
  {Valentino}}\ and\ \bibinfo {author} {\bibfnamefont {L.}~\bibnamefont
  {Mersini-Houghton}},\ }\href@noop {} {\enquote {\bibinfo {title} {Testing
  predictions of the quantum landscape multiverse 3: The hilltop inflationary
  potential},}\ } (\bibinfo {year} {2018}),\ \Eprint
  {http://arxiv.org/abs/1807.10833} {arXiv:1807.10833 [astro-ph.CO]}
  \BibitemShut {NoStop}%
\bibitem [{\citenamefont {Bouhmadi-L{\'{o}}pez}\ \emph
  {et~al.}(2019)\citenamefont {Bouhmadi-L{\'{o}}pez}, \citenamefont {Kr\"amer},
  \citenamefont {Morais},\ and\ \citenamefont {Robles-P{\'{e}}rez}}]{3}%
  \BibitemOpen
  \bibfield  {author} {\bibinfo {author} {\bibfnamefont {M.}~\bibnamefont
  {Bouhmadi-L{\'{o}}pez}}, \bibinfo {author} {\bibfnamefont {M.}~\bibnamefont
  {Kr\"amer}}, \bibinfo {author} {\bibfnamefont {J.}~\bibnamefont {Morais}}, \
  and\ \bibinfo {author} {\bibfnamefont {S.}~\bibnamefont
  {Robles-P{\'{e}}rez}},\ }\href {\doibase 10.1088/1475-7516/2019/02/057}
  {\bibfield  {journal} {\bibinfo  {journal} {Journal of Cosmology and
  Astroparticle Physics}\ }\textbf {\bibinfo {volume} {2019}},\ \bibinfo
  {pages} {057} (\bibinfo {year} {2019})}\BibitemShut {NoStop}%
\bibitem [{\citenamefont {Mersini-Houghton}\ and\ \citenamefont
  {Holman}(2009)}]{4}%
  \BibitemOpen
  \bibfield  {author} {\bibinfo {author} {\bibfnamefont {L.}~\bibnamefont
  {Mersini-Houghton}}\ and\ \bibinfo {author} {\bibfnamefont {R.}~\bibnamefont
  {Holman}},\ }\href {\doibase 10.1088/1475-7516/2009/02/006} {\bibfield
  {journal} {\bibinfo  {journal} {Journal of Cosmology and Astroparticle
  Physics}\ }\textbf {\bibinfo {volume} {2009}},\ \bibinfo {pages} {006}
  (\bibinfo {year} {2009})}\BibitemShut {NoStop}%
\bibitem [{\citenamefont {Mersini-Houghton}(2017)}]{5}%
  \BibitemOpen
  \bibfield  {author} {\bibinfo {author} {\bibfnamefont {L.}~\bibnamefont
  {Mersini-Houghton}},\ }\href {\doibase 10.1088/1361-6382/34/4/047001}
  {\bibfield  {journal} {\bibinfo  {journal} {Classical and Quantum Gravity}\
  }\textbf {\bibinfo {volume} {34}},\ \bibinfo {pages} {047001} (\bibinfo
  {year} {2017})}\BibitemShut {NoStop}%
\bibitem [{\citenamefont {Hartle}\ and\ \citenamefont
  {Hawking}(1983)}]{Hartle1983}%
  \BibitemOpen
  \bibfield  {author} {\bibinfo {author} {\bibfnamefont {J.~B.}\ \bibnamefont
  {Hartle}}\ and\ \bibinfo {author} {\bibfnamefont {S.~W.}\ \bibnamefont
  {Hawking}},\ }\href@noop {} {\bibfield  {journal} {\bibinfo  {journal} {Phys.
  Rev. D}\ }\textbf {\bibinfo {volume} {28}},\ \bibinfo {pages} {2960}
  (\bibinfo {year} {1983})}\BibitemShut {NoStop}%
\bibitem [{\citenamefont {Vilenkin}(1988)}]{Vilenkin1988}%
  \BibitemOpen
  \bibfield  {author} {\bibinfo {author} {\bibfnamefont {A.}~\bibnamefont
  {Vilenkin}},\ }\href@noop {} {\bibfield  {journal} {\bibinfo  {journal}
  {Phys. Rev. D}\ }\textbf {\bibinfo {volume} {37}},\ \bibinfo {pages} {888}
  (\bibinfo {year} {1988})}\BibitemShut {NoStop}%
\bibitem [{\citenamefont {Kiefer}(2013)}]{Kiefer2013}%
  \BibitemOpen
  \bibfield  {author} {\bibinfo {author} {\bibfnamefont {C.}~\bibnamefont
  {Kiefer}},\ }\href@noop {} {\bibfield  {journal} {\bibinfo  {journal} {Math.
  Phys.}\ }\textbf {\bibinfo {volume} {2013}},\ \bibinfo {pages} {509316}
  (\bibinfo {year} {2013})},\ \Eprint {http://arxiv.org/abs/arXiv:1401.3578}
  {arXiv:1401.3578} \BibitemShut {NoStop}%
\bibitem [{\citenamefont {Caderni}\ and\ \citenamefont
  {Martellini}(1984)}]{Caderni1984}%
  \BibitemOpen
  \bibfield  {author} {\bibinfo {author} {\bibfnamefont {N.}~\bibnamefont
  {Caderni}}\ and\ \bibinfo {author} {\bibfnamefont {M.}~\bibnamefont
  {Martellini}},\ }\href@noop {} {\bibfield  {journal} {\bibinfo  {journal}
  {Int. J. Theor. Phys.}\ }\textbf {\bibinfo {volume} {23}},\ \bibinfo {pages}
  {233} (\bibinfo {year} {1984})}\BibitemShut {NoStop}%
\bibitem [{\citenamefont {McGuigan}(1988)}]{McGuigan1988}%
  \BibitemOpen
  \bibfield  {author} {\bibinfo {author} {\bibfnamefont {M.}~\bibnamefont
  {McGuigan}},\ }\href@noop {} {\bibfield  {journal} {\bibinfo  {journal}
  {Phys. Rev. D}\ }\textbf {\bibinfo {volume} {38}},\ \bibinfo {pages} {3031}
  (\bibinfo {year} {1988})}\BibitemShut {NoStop}%
\bibitem [{\citenamefont {De~Witt}(1967)}]{DeWitt1967}%
  \BibitemOpen
  \bibfield  {author} {\bibinfo {author} {\bibfnamefont {B.~S.}\ \bibnamefont
  {De~Witt}},\ }\href@noop {} {\bibfield  {journal} {\bibinfo  {journal} {Phys.
  Rev.}\ }\textbf {\bibinfo {volume} {160}},\ \bibinfo {pages} {1113} (\bibinfo
  {year} {1967})}\BibitemShut {NoStop}%
\bibitem [{\citenamefont {Robles-P\'erez}\ \emph {et~al.}(2017)\citenamefont
  {Robles-P\'erez}, \citenamefont {Balcerzak}, \citenamefont {D\k{a}browski},\
  and\ \citenamefont {Kr\"amer}}]{Interuniversal}%
  \BibitemOpen
  \bibfield  {author} {\bibinfo {author} {\bibfnamefont {S.}~\bibnamefont
  {Robles-P\'erez}}, \bibinfo {author} {\bibfnamefont {A.}~\bibnamefont
  {Balcerzak}}, \bibinfo {author} {\bibfnamefont {M.~P.}\ \bibnamefont
  {D\k{a}browski}}, \ and\ \bibinfo {author} {\bibfnamefont {M.}~\bibnamefont
  {Kr\"amer}},\ }\href {\doibase 10.1103/PhysRevD.95.083505} {\bibfield
  {journal} {\bibinfo  {journal} {Phys. Rev. D}\ }\textbf {\bibinfo {volume}
  {95}},\ \bibinfo {pages} {083505} (\bibinfo {year} {2017})}\BibitemShut
  {NoStop}%
\bibitem [{\citenamefont {Lewis}\ and\ \citenamefont
  {Riesenfeld}(1969)}]{LewisRiesenfeld}%
  \BibitemOpen
  \bibfield  {author} {\bibinfo {author} {\bibfnamefont {H.~R.}\ \bibnamefont
  {Lewis}}\ and\ \bibinfo {author} {\bibfnamefont {W.~B.}\ \bibnamefont
  {Riesenfeld}},\ }\href {\doibase 10.1063/1.1664991} {\bibfield  {journal}
  {\bibinfo  {journal} {Journal of Mathematical Physics}\ }\textbf {\bibinfo
  {volume} {10}},\ \bibinfo {pages} {1458} (\bibinfo {year}
  {1969})}\BibitemShut {NoStop}%
\bibitem [{\citenamefont {Kim}\ \emph {et~al.}(1996)\citenamefont {Kim},
  \citenamefont {Lee}, \citenamefont {Ji},\ and\ \citenamefont
  {Kim}}]{KimLewis}%
  \BibitemOpen
  \bibfield  {author} {\bibinfo {author} {\bibfnamefont {H.-C.}\ \bibnamefont
  {Kim}}, \bibinfo {author} {\bibfnamefont {M.-H.}\ \bibnamefont {Lee}},
  \bibinfo {author} {\bibfnamefont {J.-Y.}\ \bibnamefont {Ji}}, \ and\ \bibinfo
  {author} {\bibfnamefont {J.~K.}\ \bibnamefont {Kim}},\ }\href@noop {}
  {\bibfield  {journal} {\bibinfo  {journal} {Phys. Rev. A}\ }\textbf {\bibinfo
  {volume} {53}},\ \bibinfo {pages} {3767} (\bibinfo {year}
  {1996})}\BibitemShut {NoStop}%
\bibitem [{\citenamefont {Kiefer}(1988)}]{PacketsKiefer}%
  \BibitemOpen
  \bibfield  {author} {\bibinfo {author} {\bibfnamefont {C.}~\bibnamefont
  {Kiefer}},\ }\href {\doibase 10.1103/PhysRevD.38.1761} {\bibfield  {journal}
  {\bibinfo  {journal} {Phys. Rev. D}\ }\textbf {\bibinfo {volume} {38}},\
  \bibinfo {pages} {1761} (\bibinfo {year} {1988})}\BibitemShut {NoStop}%
\bibitem [{\citenamefont {Robles-P{\'e}rez}(2019)}]{RP2019c}%
  \BibitemOpen
  \bibfield  {author} {\bibinfo {author} {\bibfnamefont {S.~J.}\ \bibnamefont
  {Robles-P{\'e}rez}},\ }\href@noop {} {\bibfield  {journal} {\bibinfo
  {journal} {Universe}\ }\textbf {\bibinfo {volume} {5}},\ \bibinfo {pages}
  {150} (\bibinfo {year} {2019})},\ \Eprint
  {http://arxiv.org/abs/arXiv:1901.03387} {arXiv:1901.03387} \BibitemShut
  {NoStop}%
\bibitem [{\citenamefont {Robles-P{\'e}rez}(2020)}]{RP2020a}%
  \BibitemOpen
  \bibfield  {author} {\bibinfo {author} {\bibfnamefont {S.~J.}\ \bibnamefont
  {Robles-P{\'e}rez}},\ }\href@noop {} {\bibfield  {journal} {\bibinfo
  {journal} {Acta Phys. Pol.}\ }\textbf {\bibinfo {volume} {13}},\ \bibinfo
  {pages} {325} (\bibinfo {year} {2020})}\BibitemShut {NoStop}%
\bibitem [{\citenamefont {Birrell}\ and\ \citenamefont
  {Davies}(1984)}]{Birrell:1982ix}%
  \BibitemOpen
  \bibfield  {author} {\bibinfo {author} {\bibfnamefont {N.~D.}\ \bibnamefont
  {Birrell}}\ and\ \bibinfo {author} {\bibfnamefont {P.~C.~W.}\ \bibnamefont
  {Davies}},\ }\href {\doibase 10.1017/CBO9780511622632} {\emph {\bibinfo
  {title} {{Quantum Fields in Curved Space}}}},\ Cambridge Monographs on
  Mathematical Physics\ (\bibinfo  {publisher} {Cambridge Univ. Press},\
  \bibinfo {address} {Cambridge, UK},\ \bibinfo {year} {1984})\BibitemShut
  {NoStop}%
\bibitem [{\citenamefont {Tsallis}(1988)}]{Tsallis}%
  \BibitemOpen
  \bibfield  {author} {\bibinfo {author} {\bibfnamefont {C.}~\bibnamefont
  {Tsallis}},\ }\href {\doibase 10.1007/BF01016429} {\bibfield  {journal}
  {\bibinfo  {journal} {Journal of Statistical Physics}\ }\textbf {\bibinfo
  {volume} {52}},\ \bibinfo {pages} {479} (\bibinfo {year} {1988})}\BibitemShut
  {NoStop}%
\bibitem [{\citenamefont {Tsallis}(2009)}]{TSALLIS2009}%
  \BibitemOpen
  \bibfield  {author} {\bibinfo {author} {\bibfnamefont {C.}~\bibnamefont
  {Tsallis}},\ }\href
  {http://www.scielo.br/scielo.php?script=sci_arttext&pid=S0103-97332009000400002&nrm=iso}
  {\bibfield  {journal} {\bibinfo  {journal} {{Brazilian Journal of Physics}}\
  }\textbf {\bibinfo {volume} {39}},\ \bibinfo {pages} {337 } (\bibinfo {year}
  {2009})}\BibitemShut {NoStop}%
\bibitem [{\citenamefont {{R\'enyi}}(1959)}]{zbMATH03143975}%
  \BibitemOpen
  \bibfield  {author} {\bibinfo {author} {\bibfnamefont {A.}~\bibnamefont
  {{R\'enyi}}},\ }\href@noop {} {\bibfield  {journal} {\bibinfo  {journal}
  {{Acta Math. Acad. Sci. Hung.}}\ }\textbf {\bibinfo {volume} {10}},\ \bibinfo
  {pages} {193} (\bibinfo {year} {1959})}\BibitemShut {NoStop}%
\bibitem [{\citenamefont {R{\'e}nyi}(1961)}]{Renyi}%
  \BibitemOpen
  \bibfield  {author} {\bibinfo {author} {\bibfnamefont {A.}~\bibnamefont
  {R{\'e}nyi}},\ }in\ \href@noop {} {\emph {\bibinfo {booktitle} {Proceedings
  of the Fourth Berkeley Symposium on Mathematical Statistics and Probability,
  Volume 1: Contributions to the Theory of Statistics}}}\ (\bibinfo
  {publisher} {University of California Press},\ \bibinfo {address} {Berkeley,
  Calif.},\ \bibinfo {year} {1961})\ pp.\ \bibinfo {pages}
  {547--561}\BibitemShut {NoStop}%
\bibitem [{\citenamefont {Plenio}\ and\ \citenamefont
  {Vedral}(1998)}]{Plenio1998}%
  \BibitemOpen
  \bibfield  {author} {\bibinfo {author} {\bibfnamefont {M.~B.}\ \bibnamefont
  {Plenio}}\ and\ \bibinfo {author} {\bibfnamefont {V.}~\bibnamefont
  {Vedral}},\ }\href@noop {} {\bibfield  {journal} {\bibinfo  {journal}
  {Comtemp. Phys.}\ }\textbf {\bibinfo {volume} {39}},\ \bibinfo {pages} {431}
  (\bibinfo {year} {1998})}\BibitemShut {NoStop}%
\bibitem [{\citenamefont {D\k{a}browski}(1996)}]{DABROWSKI1996}%
  \BibitemOpen
  \bibfield  {author} {\bibinfo {author} {\bibfnamefont {M.~P.}\ \bibnamefont
  {D\k{a}browski}},\ }\href {\doibase https://doi.org/10.1006/aphy.1996.0057}
  {\bibfield  {journal} {\bibinfo  {journal} {Annals of Physics}\ }\textbf
  {\bibinfo {volume} {248}},\ \bibinfo {pages} {199 } (\bibinfo {year}
  {1996})}\BibitemShut {NoStop}%
\bibitem [{\citenamefont {Graham}\ \emph
  {et~al.}(2014{\natexlab{a}})\citenamefont {Graham}, \citenamefont {Horn},
  \citenamefont {Kachru}, \citenamefont {Rajendran},\ and\ \citenamefont
  {Torroba}}]{Graham:2011nb}%
  \BibitemOpen
  \bibfield  {author} {\bibinfo {author} {\bibfnamefont {P.~W.}\ \bibnamefont
  {Graham}}, \bibinfo {author} {\bibfnamefont {B.}~\bibnamefont {Horn}},
  \bibinfo {author} {\bibfnamefont {S.}~\bibnamefont {Kachru}}, \bibinfo
  {author} {\bibfnamefont {S.}~\bibnamefont {Rajendran}}, \ and\ \bibinfo
  {author} {\bibfnamefont {G.}~\bibnamefont {Torroba}},\ }\href {\doibase
  10.1007/JHEP02(2014)029} {\bibfield  {journal} {\bibinfo  {journal} {JHEP}\
  }\textbf {\bibinfo {volume} {02}},\ \bibinfo {pages} {029} (\bibinfo {year}
  {2014}{\natexlab{a}})},\ \Eprint {http://arxiv.org/abs/1109.0282}
  {arXiv:1109.0282 [hep-th]} \BibitemShut {NoStop}%
\bibitem [{\citenamefont {Mithani}\ and\ \citenamefont
  {Vilenkin}(2012)}]{Mithani:2011en}%
  \BibitemOpen
  \bibfield  {author} {\bibinfo {author} {\bibfnamefont {A.~T.}\ \bibnamefont
  {Mithani}}\ and\ \bibinfo {author} {\bibfnamefont {A.}~\bibnamefont
  {Vilenkin}},\ }\href {\doibase 10.1088/1475-7516/2012/01/028} {\bibfield
  {journal} {\bibinfo  {journal} {JCAP}\ }\textbf {\bibinfo {volume} {1201}},\
  \bibinfo {pages} {028} (\bibinfo {year} {2012})},\ \Eprint
  {http://arxiv.org/abs/1110.4096} {arXiv:1110.4096 [hep-th]} \BibitemShut
  {NoStop}%
\bibitem [{\citenamefont {Graham}\ \emph
  {et~al.}(2014{\natexlab{b}})\citenamefont {Graham}, \citenamefont {Horn},
  \citenamefont {Rajendran},\ and\ \citenamefont {Torroba}}]{Graham:2014pca}%
  \BibitemOpen
  \bibfield  {author} {\bibinfo {author} {\bibfnamefont {P.~W.}\ \bibnamefont
  {Graham}}, \bibinfo {author} {\bibfnamefont {B.}~\bibnamefont {Horn}},
  \bibinfo {author} {\bibfnamefont {S.}~\bibnamefont {Rajendran}}, \ and\
  \bibinfo {author} {\bibfnamefont {G.}~\bibnamefont {Torroba}},\ }\href
  {\doibase 10.1007/JHEP08(2014)163} {\bibfield  {journal} {\bibinfo  {journal}
  {JHEP}\ }\textbf {\bibinfo {volume} {08}},\ \bibinfo {pages} {163} (\bibinfo
  {year} {2014}{\natexlab{b}})},\ \Eprint {http://arxiv.org/abs/1405.0282}
  {arXiv:1405.0282 [hep-th]} \BibitemShut {NoStop}%
\bibitem [{\citenamefont {Mithani}\ and\ \citenamefont
  {Vilenkin}(2014)}]{Mithani:2014jva}%
  \BibitemOpen
  \bibfield  {author} {\bibinfo {author} {\bibfnamefont {A.~T.}\ \bibnamefont
  {Mithani}}\ and\ \bibinfo {author} {\bibfnamefont {A.}~\bibnamefont
  {Vilenkin}},\ }\href {\doibase 10.1088/1475-7516/2014/05/006} {\bibfield
  {journal} {\bibinfo  {journal} {JCAP}\ }\textbf {\bibinfo {volume} {1405}},\
  \bibinfo {pages} {006} (\bibinfo {year} {2014})},\ \Eprint
  {http://arxiv.org/abs/1403.0818} {arXiv:1403.0818 [hep-th]} \BibitemShut
  {NoStop}%
\bibitem [{\citenamefont {Mithani}\ and\ \citenamefont
  {Vilenkin}(2015)}]{Mithani:2014toa}%
  \BibitemOpen
  \bibfield  {author} {\bibinfo {author} {\bibfnamefont {A.~T.}\ \bibnamefont
  {Mithani}}\ and\ \bibinfo {author} {\bibfnamefont {A.}~\bibnamefont
  {Vilenkin}},\ }\href {\doibase 10.1088/1475-7516/2015/07/010} {\bibfield
  {journal} {\bibinfo  {journal} {JCAP}\ }\textbf {\bibinfo {volume} {1507}},\
  \bibinfo {pages} {010} (\bibinfo {year} {2015})},\ \Eprint
  {http://arxiv.org/abs/1407.5361} {arXiv:1407.5361 [hep-th]} \BibitemShut
  {NoStop}%
\bibitem [{\citenamefont {D\k{a}browski}\ and\ \citenamefont
  {Larsen}(1995)}]{Dabrowski:1995jt}%
  \BibitemOpen
  \bibfield  {author} {\bibinfo {author} {\bibfnamefont {M.~P.}\ \bibnamefont
  {D\k{a}browski}}\ and\ \bibinfo {author} {\bibfnamefont {A.~L.}\ \bibnamefont
  {Larsen}},\ }\href {\doibase 10.1103/PhysRevD.52.3424} {\bibfield  {journal}
  {\bibinfo  {journal} {Phys. Rev.}\ }\textbf {\bibinfo {volume} {D52}},\
  \bibinfo {pages} {3424} (\bibinfo {year} {1995})}\BibitemShut {NoStop}%
\bibitem [{\citenamefont {Damour}\ and\ \citenamefont
  {Vilenkin}(2019)}]{PhysRevD.100.083525}%
  \BibitemOpen
  \bibfield  {author} {\bibinfo {author} {\bibfnamefont {T.}~\bibnamefont
  {Damour}}\ and\ \bibinfo {author} {\bibfnamefont {A.}~\bibnamefont
  {Vilenkin}},\ }\href {\doibase 10.1103/PhysRevD.100.083525} {\bibfield
  {journal} {\bibinfo  {journal} {Phys. Rev. D}\ }\textbf {\bibinfo {volume}
  {100}},\ \bibinfo {pages} {083525} (\bibinfo {year} {2019})}\BibitemShut
  {NoStop}%
\bibitem [{\citenamefont {Caldwell}(2002)}]{Caldwell:1999ew}%
  \BibitemOpen
  \bibfield  {author} {\bibinfo {author} {\bibfnamefont {R.}~\bibnamefont
  {Caldwell}},\ }\href {\doibase 10.1016/S0370-2693(02)02589-3} {\bibfield
  {journal} {\bibinfo  {journal} {Phys. Lett. B}\ }\textbf {\bibinfo {volume}
  {545}},\ \bibinfo {pages} {23} (\bibinfo {year} {2002})},\ \Eprint
  {http://arxiv.org/abs/astro-ph/9908168} {arXiv:astro-ph/9908168} \BibitemShut
  {NoStop}%
\bibitem [{\citenamefont {D\k{a}browski}\ \emph {et~al.}(2003)\citenamefont
  {D\k{a}browski}, \citenamefont {Stachowiak},\ and\ \citenamefont
  {Szydlowski}}]{Dabrowski:2003jm}%
  \BibitemOpen
  \bibfield  {author} {\bibinfo {author} {\bibfnamefont {M.~P.}\ \bibnamefont
  {D\k{a}browski}}, \bibinfo {author} {\bibfnamefont {T.}~\bibnamefont
  {Stachowiak}}, \ and\ \bibinfo {author} {\bibfnamefont {M.}~\bibnamefont
  {Szydlowski}},\ }\href {\doibase 10.1103/PhysRevD.68.103519} {\bibfield
  {journal} {\bibinfo  {journal} {Phys. Rev. D}\ }\textbf {\bibinfo {volume}
  {68}},\ \bibinfo {pages} {103519} (\bibinfo {year} {2003})},\ \Eprint
  {http://arxiv.org/abs/hep-th/0307128} {arXiv:hep-th/0307128} \BibitemShut
  {NoStop}%
\bibitem [{\citenamefont {Barrow}(2004)}]{Barrow:2004xh}%
  \BibitemOpen
  \bibfield  {author} {\bibinfo {author} {\bibfnamefont {J.~D.}\ \bibnamefont
  {Barrow}},\ }\href {\doibase 10.1088/0264-9381/21/11/L03} {\bibfield
  {journal} {\bibinfo  {journal} {Class. Quant. Grav.}\ }\textbf {\bibinfo
  {volume} {21}},\ \bibinfo {pages} {L79} (\bibinfo {year} {2004})},\ \Eprint
  {http://arxiv.org/abs/gr-qc/0403084} {arXiv:gr-qc/0403084} \BibitemShut
  {NoStop}%
\bibitem [{\citenamefont {Kamenshchik}\ \emph {et~al.}(2007)\citenamefont
  {Kamenshchik}, \citenamefont {Kiefer},\ and\ \citenamefont
  {Sandh\"ofer}}]{Exotic2007}%
  \BibitemOpen
  \bibfield  {author} {\bibinfo {author} {\bibfnamefont {A.~Y.}\ \bibnamefont
  {Kamenshchik}}, \bibinfo {author} {\bibfnamefont {C.}~\bibnamefont {Kiefer}},
  \ and\ \bibinfo {author} {\bibfnamefont {B.}~\bibnamefont {Sandh\"ofer}},\
  }\href {\doibase 10.1103/PhysRevD.76.064032} {\bibfield  {journal} {\bibinfo
  {journal} {Phys. Rev. D}\ }\textbf {\bibinfo {volume} {76}},\ \bibinfo
  {pages} {064032} (\bibinfo {year} {2007})}\BibitemShut {NoStop}%
\bibitem [{\citenamefont {Nojiri}\ \emph {et~al.}(2005)\citenamefont {Nojiri},
  \citenamefont {Odintsov},\ and\ \citenamefont {Tsujikawa}}]{Nojiri:2005sx}%
  \BibitemOpen
  \bibfield  {author} {\bibinfo {author} {\bibfnamefont {S.}~\bibnamefont
  {Nojiri}}, \bibinfo {author} {\bibfnamefont {S.~D.}\ \bibnamefont
  {Odintsov}}, \ and\ \bibinfo {author} {\bibfnamefont {S.}~\bibnamefont
  {Tsujikawa}},\ }\href {\doibase 10.1103/PhysRevD.71.063004} {\bibfield
  {journal} {\bibinfo  {journal} {Phys. Rev. D}\ }\textbf {\bibinfo {volume}
  {71}},\ \bibinfo {pages} {063004} (\bibinfo {year} {2005})},\ \Eprint
  {http://arxiv.org/abs/hep-th/0501025} {arXiv:hep-th/0501025} \BibitemShut
  {NoStop}%
\bibitem [{\citenamefont {D\k{a}browski}\ and\ \citenamefont
  {Denkiewicz}(2009)}]{Dabrowski:2009kg}%
  \BibitemOpen
  \bibfield  {author} {\bibinfo {author} {\bibfnamefont {M.~P.}\ \bibnamefont
  {D\k{a}browski}}\ and\ \bibinfo {author} {\bibfnamefont {T.}~\bibnamefont
  {Denkiewicz}},\ }\href {\doibase 10.1103/PhysRevD.79.063521} {\bibfield
  {journal} {\bibinfo  {journal} {Phys. Rev. D}\ }\textbf {\bibinfo {volume}
  {79}},\ \bibinfo {pages} {063521} (\bibinfo {year} {2009})},\ \Eprint
  {http://arxiv.org/abs/0902.3107} {arXiv:0902.3107 [gr-qc]} \BibitemShut
  {NoStop}%
\bibitem [{\citenamefont {Frampton}\ \emph {et~al.}(2011)\citenamefont
  {Frampton}, \citenamefont {Ludwick},\ and\ \citenamefont
  {Scherrer}}]{Frampton:2011sp}%
  \BibitemOpen
  \bibfield  {author} {\bibinfo {author} {\bibfnamefont {P.~H.}\ \bibnamefont
  {Frampton}}, \bibinfo {author} {\bibfnamefont {K.~J.}\ \bibnamefont
  {Ludwick}}, \ and\ \bibinfo {author} {\bibfnamefont {R.~J.}\ \bibnamefont
  {Scherrer}},\ }\href {\doibase 10.1103/PhysRevD.84.063003} {\bibfield
  {journal} {\bibinfo  {journal} {Phys. Rev. D}\ }\textbf {\bibinfo {volume}
  {84}},\ \bibinfo {pages} {063003} (\bibinfo {year} {2011})},\ \Eprint
  {http://arxiv.org/abs/1106.4996} {arXiv:1106.4996 [astro-ph.CO]} \BibitemShut
  {NoStop}%
\bibitem [{\citenamefont {D\k{a}browski}\ and\ \citenamefont
  {Denkiewicz}(2010)}]{Dabrowski:2009pc}%
  \BibitemOpen
  \bibfield  {author} {\bibinfo {author} {\bibfnamefont {M.~P.}\ \bibnamefont
  {D\k{a}browski}}\ and\ \bibinfo {author} {\bibfnamefont {T.}~\bibnamefont
  {Denkiewicz}},\ }\href {\doibase 10.1063/1.3462686} {\bibfield  {journal}
  {\bibinfo  {journal} {AIP Conf. Proc.}\ }\textbf {\bibinfo {volume} {1241}},\
  \bibinfo {pages} {561} (\bibinfo {year} {2010})},\ \Eprint
  {http://arxiv.org/abs/0910.0023} {arXiv:0910.0023 [gr-qc]} \BibitemShut
  {NoStop}%
\bibitem [{\citenamefont {D\k{a}browski}(2014)}]{Dabrowski:2014fha}%
  \BibitemOpen
  \bibfield  {author} {\bibinfo {author} {\bibfnamefont {M.~P.}\ \bibnamefont
  {D\k{a}browski}},\ }\enquote {\bibinfo {title} {{Are singularities the limits
  of cosmology?}}}\ in\ \href@noop {} {\emph {\bibinfo {booktitle}
  {{Mathematical Structures of the Universe}}}},\ \bibinfo {editor} {edited by\
  \bibinfo {editor} {\bibfnamefont {M.}~\bibnamefont {Heller}}, \bibinfo
  {editor} {\bibfnamefont {M.}~\bibnamefont {Eckstein}}, \ and\ \bibinfo
  {editor} {\bibfnamefont {S.}~\bibnamefont {Szybka}}}\ (\bibinfo {year}
  {2014})\ pp.\ \bibinfo {pages} {101--118},\ \Eprint
  {http://arxiv.org/abs/1407.4851} {arXiv:1407.4851 [gr-qc]} \BibitemShut
  {NoStop}%
\bibitem [{\citenamefont {D\k{a}browski}\ and\ \citenamefont
  {Marosek}(2018)}]{Dabrowski:2018ucy}%
  \BibitemOpen
  \bibfield  {author} {\bibinfo {author} {\bibfnamefont {M.~P.}\ \bibnamefont
  {D\k{a}browski}}\ and\ \bibinfo {author} {\bibfnamefont {K.}~\bibnamefont
  {Marosek}},\ }\href {\doibase 10.1007/s10714-018-2482-1} {\bibfield
  {journal} {\bibinfo  {journal} {Gen. Rel. Grav.}\ }\textbf {\bibinfo {volume}
  {50}},\ \bibinfo {pages} {160} (\bibinfo {year} {2018})},\ \Eprint
  {http://arxiv.org/abs/1806.00601} {arXiv:1806.00601 [gr-qc]} \BibitemShut
  {NoStop}%
\bibitem [{\citenamefont {Tipler}(1977)}]{TIPLER19778}%
  \BibitemOpen
  \bibfield  {author} {\bibinfo {author} {\bibfnamefont {F.~J.}\ \bibnamefont
  {Tipler}},\ }\href {\doibase https://doi.org/10.1016/0375-9601(77)90508-4}
  {\bibfield  {journal} {\bibinfo  {journal} {Physics Letters A}\ }\textbf
  {\bibinfo {volume} {64}},\ \bibinfo {pages} {8 } (\bibinfo {year}
  {1977})}\BibitemShut {NoStop}%
\bibitem [{\citenamefont {Krolak}(1986)}]{Krolak_1986}%
  \BibitemOpen
  \bibfield  {author} {\bibinfo {author} {\bibfnamefont {A.}~\bibnamefont
  {Krolak}},\ }\href {\doibase 10.1088/0264-9381/3/3/004} {\bibfield  {journal}
  {\bibinfo  {journal} {Classical and Quantum Gravity}\ }\textbf {\bibinfo
  {volume} {3}},\ \bibinfo {pages} {267} (\bibinfo {year} {1986})}\BibitemShut
  {NoStop}%
\bibitem [{\citenamefont {Bouhmadi-L\'opez}\ \emph {et~al.}(2009)\citenamefont
  {Bouhmadi-L\'opez}, \citenamefont {Kiefer}, \citenamefont {Sandh\"ofer},\
  and\ \citenamefont {Moniz}}]{Exotic2009}%
  \BibitemOpen
  \bibfield  {author} {\bibinfo {author} {\bibfnamefont {M.}~\bibnamefont
  {Bouhmadi-L\'opez}}, \bibinfo {author} {\bibfnamefont {C.}~\bibnamefont
  {Kiefer}}, \bibinfo {author} {\bibfnamefont {B.}~\bibnamefont {Sandh\"ofer}},
  \ and\ \bibinfo {author} {\bibfnamefont {P.~V.}\ \bibnamefont {Moniz}},\
  }\href {\doibase 10.1103/PhysRevD.79.124035} {\bibfield  {journal} {\bibinfo
  {journal} {Phys. Rev. D}\ }\textbf {\bibinfo {volume} {79}},\ \bibinfo
  {pages} {124035} (\bibinfo {year} {2009})}\BibitemShut {NoStop}%
\bibitem [{\citenamefont {Albarran}\ \emph {et~al.}(2016)\citenamefont
  {Albarran}, \citenamefont {Bouhmadi-L\'opez}, \citenamefont {Kiefer},
  \citenamefont {Marto},\ and\ \citenamefont {Vargas~Moniz}}]{Exotic2016}%
  \BibitemOpen
  \bibfield  {author} {\bibinfo {author} {\bibfnamefont {I.}~\bibnamefont
  {Albarran}}, \bibinfo {author} {\bibfnamefont {M.}~\bibnamefont
  {Bouhmadi-L\'opez}}, \bibinfo {author} {\bibfnamefont {C.}~\bibnamefont
  {Kiefer}}, \bibinfo {author} {\bibfnamefont {J.~a.}\ \bibnamefont {Marto}}, \
  and\ \bibinfo {author} {\bibfnamefont {P.}~\bibnamefont {Vargas~Moniz}},\
  }\href {\doibase 10.1103/PhysRevD.94.063536} {\bibfield  {journal} {\bibinfo
  {journal} {Phys. Rev. D}\ }\textbf {\bibinfo {volume} {94}},\ \bibinfo
  {pages} {063536} (\bibinfo {year} {2016})}\BibitemShut {NoStop}%
\bibitem [{\citenamefont {{Shannon}}(1948)}]{Shannon}%
  \BibitemOpen
  \bibfield  {author} {\bibinfo {author} {\bibfnamefont {C.~E.}\ \bibnamefont
  {{Shannon}}},\ }\href@noop {} {\bibfield  {journal} {\bibinfo  {journal} {The
  Bell System Technical Journal}\ }\textbf {\bibinfo {volume} {27}},\ \bibinfo
  {pages} {623} (\bibinfo {year} {1948})}\BibitemShut {NoStop}%
\bibitem [{\citenamefont {{\v{S}}teigl}\ and\ \citenamefont
  {Hinterleitner}(2006)}]{TeiglFactor}%
  \BibitemOpen
  \bibfield  {author} {\bibinfo {author} {\bibfnamefont {R.}~\bibnamefont
  {{\v{S}}teigl}}\ and\ \bibinfo {author} {\bibfnamefont {F.}~\bibnamefont
  {Hinterleitner}},\ }\href {\doibase 10.1088/0264-9381/23/11/013} {\bibfield
  {journal} {\bibinfo  {journal} {Classical and Quantum Gravity}\ }\textbf
  {\bibinfo {volume} {23}},\ \bibinfo {pages} {3879} (\bibinfo {year}
  {2006})}\BibitemShut {NoStop}%
\end{thebibliography}%

\end{document}